\documentclass[a4paper,11pt]{article}

\usepackage{multicol}
\usepackage{graphicx} 
\usepackage{bbold}
\usepackage{mathtools}
\usepackage{amsmath}
\usepackage{amsfonts}
\usepackage{amssymb}
\usepackage{slashed} 
\usepackage{color}
\usepackage[table]{xcolor}
\usepackage{tabularx}
\usepackage{tablestyles}  
\usepackage{hyperref}
\usepackage{bm}
\usepackage{mathrsfs}
\usepackage{cite}
\usepackage{physics}
\usepackage{bbding}
\usepackage[table]{xcolor}
\usepackage{booktabs}

\def\eq#1{{Eq.~(\ref{#1})}}
\def\eqs#1#2{{Eqs.~(\ref{#1})--(\ref{#2})}}

\def\abs#1{\left| #1\right|}

\def\Tr{\mbox{Tr}\,}

\def\di{\mbox{d}}

\colorlet{grayline}{gray!70}
\definecolor{mf}{rgb}{204,0,0}
\definecolor{blueline}{rgb}{0,0.27,0.55}
\definecolor{DarkGray}{gray}{0.4}
\definecolor{Gray}{gray}{0.6}
\definecolor{oucrimsonred}{rgb}{0.6, 0.0, 0.0}
\definecolor{persianblue}{rgb}{0.11, 0.22, 0.73}
\definecolor{forestgreen}{rgb}{0.13,0.35,0.13}
 \hypersetup{colorlinks, citecolor=forestgreen, linkcolor=forestgreen, urlcolor=forestgreen}
\newcommand{\be}{\begin{equation}}
\newcommand{\ee}{\end{equation}}
\newcommand{\bea}{\begin{eqnarray}}
\newcommand{\eea}{\end{eqnarray}}
\newcommand{\nn}{\nonumber}

\newcommand*\xbar[1]{%
  \hbox{\;%
    \vbox{%
      \hrule height 0.5pt 
      \kern0.5ex
      \hbox{%
        \kern-0.25em
        \ensuremath{#1}%
        \kern-0.07em
      }%
    }%
  }%
} 
\newcommand{\com}[1]{}
\newcommand{\gsim}{\lower.7ex\hbox{$\;\stackrel{\textstyle>}{\sim}\;$}}
\newcommand{\lsim}{\lower.7ex\hbox{$\;\stackrel{\textstyle<}{\sim}\;$}} 

\newcommand{\bc}{\begin{center}}
\newcommand{\ec}{\end{center}}

\newcommand{\lambdap}{\lambda^{\prime}}

\newcommand{\Bell}{{\cal B}}
\newcommand{\II}{{\cal I}_3}
\newcommand{\Aee}{A^{\ell\bar \ell}}
\newcommand{\htee}{\tilde{h}^{\ell\bar \ell}}

\newcommand{\A}{\scriptscriptstyle{A}}
\newcommand{\B}{\scriptscriptstyle{B}}
\font\beeg=cmr17 scaled 1800
\newbox\ibox
\def\versal#1{\setbox\ibox=\hbox{{\beeg #1}~}
	    \noindent\global\hangindent=\wd\ibox\global\hangafter-2
	    \sc\smash{\llap {\lower 14pt \box\ibox}}}

\newcommand{\mVV}{m_{\scriptscriptstyle{V\!V}}}

\newcommand{\G}{\scriptscriptstyle{G}}

\begin{document}

\thispagestyle{empty}
\begin{center}
  {\color{oucrimsonred}\Large {\bf Particle Collisions \& Quantum Entanglement\\
       \vspace{0.3cm} in High-Energy Collisions}}

\vspace*{1.5cm}
{\color{black}
{\bf Emidio Gabrielli$^{{a,b,c}}$}
}\\

\vspace{0.5cm}
       {\small
           {\it \color{black}
    (a) Physics Department, University of Trieste, Strada Costiera 11, \\ I-34151 Trieste, Italy}
  \\[1mm]  
{\it  \color{black}
(b) INFN, Sezione di Trieste, Via Valerio 2, I-34127 Trieste, Italy}
    \\[1mm]
  {\it \color{black}
(c) Laboratory of High-Energy and Computational Physics, NICPB, R\"avala pst 10, \\ 10143 Tallinn, Estonia}
}

\ec

 \vskip0.5cm
\bc
{\color{black}
\rule{0.7\textwidth}{0.5pt}}
\ec
\vskip1cm
\bc
{\bf ABSTRACT}
\ec

\noindent
The exploration of fundamental quantum phenomena, such as entanglement and Bell inequality violations--extensively studied in low-energy regimes--has recently extended to high-energy particle collisions. Experimentally, Bell inequality violations, which challenge Einstein's principle of local realism, were first observed in low-energy entangled photon systems by A. Aspect, J. F. Clauser, and A. Zeilinger, earning them the 2022 Nobel Prize in Physics. Particle colliders provide a novel setting for probing quantum information theory, operating at energies over ten orders of magnitude higher than previous experiments and in the presence of electroweak and strong interactions. Additionally, collider detectors offer unique advantages for quantum state reconstruction via quantum state tomography. This book chapter reviews key theoretical and experimental advancements in this emerging field, highlighting its challenges, objectives, and potential impact on both quantum information theory and high-energy physics.

\vspace*{5mm}

\vspace{5cm}
\noindent
{\small
  {\rm
Contribution to the book {\it Particle Accelerators and High-Energy Experimental Particle Physics}, edited by Theodota Lagouri, to be published by IntechOpen.}}
\newpage
\pagestyle{plain}
\tableofcontents
\newpage

\section{ Introduction}

One of the most intriguing features of quantum mechanics (QM) is the phenomenon of entanglement in composite systems. As we will show, this feature has no counterpart in classical or fully deterministic theories~\cite{Amico20081,Horodecki:2009zz,Guhne20091,Laflorencie20161}, such as hidden-variable theories (HVT)~\cite{Genovese:2005nw}, which attempt to account for the probabilistic nature of QM by postulating additional classical hidden degrees of freedom.

Deterministic theories are based on two key assumptions: {\it realism}, the notion that physical systems possess well-definite properties independent of observations, and {\it locality}, the idea that systems can be influenced only by their immediate surroundings, with no interaction propagating faster than the speed of light. The phenomenon of quantum entanglement, however, violates the principle of local realism, marking a clear boundary between classical and quantum phenomena.

At the heart of quantum entanglement lies the remarkable ability of  subsystems to remain interconnected, no matter how far apart they are, therefore revealing the non-local character of QM.
In fact, a quantum entangled system cannot be expressed as a simple combination of the states of its individual parts--thus forming a non-local unity that goes beyond classical separability. This stands in stark contrast to the concept  of local realism, which assumes that all properties of a system are predetermined and exist independently of any measurement. Importantly, the behavior of entangled systems does not violate the principles of special relativity, since no energy or information is transmitted faster than light during these measurements.

In their seminal 1935 paper, Einstein, Podolsky, and Rosen (EPR) \cite{Einstein:1935rr} supported the idea that the QM description of reality, based solely on the wave function, is incomplete if one assumes the principle of local realism. This observation opened the door to the possibility that HVTs might offer a complete and deterministic account of the quantum phenomena. The debate between the QM framework and HVTs--or between the local and non-local nature of quantum phenomena--remained unresolved until 1969, when John Bell proposed a decisive test to distinguish between them. His approach involved analyzing correlations in measurements performed independently on the spins of two spatially separated components of an entangled bi-partite system
\cite{Bell:1964,bell2004speakable}.

Under the assumption that these measurements are mutually independent and that no signal can travel faster than light between the two measurement events (i.e., locality), one can derive an upper limit on these correlations: this is what is now known as the Bell inequality \cite{Bell:1964,bell2004speakable}. QM, however, predicts stronger correlations that can exceed this classical bound, revealing its instrinsic non-local character. As a result, violations of the Bell inequality in experiments provide strong evidence against local nature of HVTs, thereby supporting the non-local property of the quantum mechanical framework~\cite{bell2004speakable,bertlmann2016quantum}.

The first experiments to test violations of the Bell inequality were carried out using entangled photons at energies on the order of a few electronvolts, see~\cite{Horodecki:2009zz} for a more complete review. In these experiments, pairs of photons were prepared in a spin-singlet state, and their polarizations (analogous to spin measurements) were measured along various directions. thus allowing to probe entanglement and potential violations of the Bell inequality.

Violations of the Bell inequality were first observed experimentally in optical setups~\cite{Aspect:1982fx,Weihs:1998gy}, by A. Aspect, J. F. Clauser, and A. Zeilinger, an achievement that earned them the 2022 Nobel Prize in Physics. 
These results sparked extensive efforts to eliminate--or at least significantly reduce--experimental loopholes.\footnote{
Loopholes are flaws in experimental setups that can weaken the conclusions of Bell tests. The most prominent is the locality loophole, which arises when the two measurement events are not space-like separated,  allowing possible information exchange that undermines the assumption of independence. Other loopholes discussed in the literature are those of {\it detection}~\cite{Pearle:1970zt}, {\it coincidence}~\cite{Larsson:2003efo}, {\it freedom of choice}~\cite{Bohm:1957zz} and {\it super determinism}~\cite{Larsson_2014}.}
Later, low-energy experiments involving photons~\cite{Hensen:2015ccp,Giustina:2015yza} and atoms~\cite{Rosenfeld:2017rka} have effectively closed nearly all significant loopholes, providing strong support for the non-local predictions of QM. Further experiments have tested the inequality~\cite{Clauser:1969ny,Clauser:1974tg}, confirming its violation even over kilometer-scale distances~\cite{Tittel:1998ja}, and also in solid-state systems~\cite{Ansmann2009ViolationOB}.

Violation of Bell inequalities is also expected in relativistic quantum field theory~\cite{Summers-Werner-1,Summers-Werner-2,Summers-Werner-3,Landau-1987,Summers-1990,Dudal:2023mij}. In particular, for a bipartite system with each party localized in space-like separated space-time regions, there always exists a state that leads to the maximal violation of a corresponding Bell inequality~\cite{Summers-Werner-1,Summers-Werner-2,Summers-Werner-3,Landau-1987,Summers-1990,Dudal:2023mij}.

Quantum entanglement and non-locality can also be investigated in high-energy physics through elementary particle collisions at colliders. In this context, entanglement manifests in spin correlations or internal quantum numbers such as flavor, since the associated symmetry groups are non-abelian~\cite{Fabbrichesi:2025aqp}. At first sight, one might assume that exploring entanglement or Bell inequality violations at colliders would require polarized beams and targets--similar to photon experiments with polarimeters--thus ruling out such studies in high-energy physics. Yet, despite such criticisms~\cite{Abel:1992kz}, this is not the case. Collider experiments typically employ unpolarized beams and detectors that do not measure particle polarizations directly. Instead, they infer the polarization of the parent particles from the angular distributions of their decay products. These distributions can be used to reconstruct correlations and translate cross-section measurements into a polarization density matrix. This method, known as {\bf quantum state tomography}, enables the measurement of entanglement and the testing of Bell non-locality at colliders, while also providing a novel tool for probing physics beyond the Standard Model (SM) through entanglement-sensitive observables~\cite{Barr:2024djo}.

At this point, a clarification is needed regarding how Bell non-locality is tested in particle physics.
The most transparent framework for testing non-locality against hidden-variable models is the black-box approach, where classical inputs can be freely chosen and the corresponding outputs recorded \cite{Eckstein:2021pgm,Altomonte:2024upf}. This is the standard method in quantum information
theory~\cite{Moroder:2013mxj,Scarani:2015vsq} as it avoids dependence on detailed knowledge of the measurement devices, which may be unavailable or subject to uncertainties. However, no existing or planned collider experiment meets the requirements for device independence. In practice, high-energy experiments probe non-locality only in a device-specific setting, which, in the language of quantum information theory, corresponds to non-locality witnesses rather than genuine Bell tests--see~\cite{Fabbrichesi:2025aqp} for more details.

Then, unlike conventional Bell tests, which rely only on joint probability measurements along selected spatial directions and thus leave the final state only partially determined (opening the door to loopholes), collider experiments allow a full reconstruction of the state via quantum tomography. This closes such loopholes and reduces Bell tests to the straightforward evaluation of suitable observables~\cite{Fabbrichesi:2025aqp}.

In particle physics, entanglement has been initially investigated using low-energy protons~\cite{Lamehi-Rachti:1976wey}, while other proposals for testing it at high-energy colliders appeared in~\cite{Tornqvist:1980af,Privitera:1991nz,Abel:1992kz}. Additional tests of entanglement and Bell inequalities have been suggested in various contexts, including positronium~\cite{Acin:2000cs,Li:2008dk}, charmonium decays~\cite{Tornqvist:1986pe,Baranov:2009zza,Baranov:2008zzb,Chen:2013epa,Qian:2020ini}, neutrino oscillations~\cite{Banerjee:2015mha}, and in the neutral mesons/anti-meson systems~\cite{Benatti2000,Bertlmann:2001ea,Banerjee:2014vga,Go:2003tx}, although only an indirect test of Bell non-locality could be probed in this last case.

Interest in the high-energy phenomenology of entanglement has been recently renewed following the demonstration that spin correlations in top-quark pairs produced at the Large Hadron Collider (LHC)\cite{Evans:2008zzb} at CERN can reveal entanglement~\cite{Afik:2020onf}, and that such systems may also exhibit violations of the Bell inequality~\cite{Fabbrichesi:2021npl}. This revival has led to a wave of studies, particularly focused on top-quark production~\cite{Severi:2021cnj,Aguilar-Saavedra:2022uye,Larkoski:2022lmv,Afik:2022kwm,Han:2023fci,Dong:2023xiw}, as well as investigations involving spin-1 mesons~\cite{Fabbrichesi:2023idl,Gabrielli:2024kbz,Fabbrichesi:2024rec}, hyperons~\cite{Gong:2021bcp,Fabbrichesi:2024rec}, tau-lepton pairs~\cite{Ehataht:2023zzt,Fabbrichesi:2024wcd}, gauge bosons originating from Higgs decays~\cite{Barr:2021zcp,Ashby-Pickering:2022umy,Aguilar-Saavedra:2022wam,Fabbrichesi:2023cev}, and vector boson scattering~\cite{Morales:2023gow}. Together, these works highlight that both entanglement and violations of the Bell inequality are accessible in a variety of processes at high-energy colliders, and that the associated phenomenology offers a novel approach to probing physics beyond the SM~\cite{Fabbrichesi:2022ovb,Maltoni:2024tul,Fabbrichesi:2023jep,Aoude:2023hxv,Bernal:2023ruk,Fabbrichesi:2024xtq,Fabbrichesi:2024wcd}. See
a recent review~\cite{Barr:2024djo} for more details.

The detection of entanglement in high-energy physics has recently been confirmed by the ATLAS~\cite{ATLAS:2023fsd} and CMS~\cite{CMS:2024pts,CMS:2024zkc} collaborations at CERN, which reported its observation with a significance exceeding $5\sigma$ through the analysis of spin correlations in top-quark pairs produced near threshold at the LHC.
In contrast, the first confirmation of Bell inequality violation at high energies was obtained by analyzing the data pertaining to the $B$ meson decays into two spin-1 mesons~\cite{Fabbrichesi:2023idl,Gabrielli:2024kbz}, based on the experimental analysis performed by the LHCb collaboration~\cite{LHCb:2013vga}. This result provides compelling evidence of Bell violation with significance well above the $5\sigma$ threshold in a bipartite system of two qutrits (three-level systems), operating at energies around $5~{\rm GeV}$. This scale is nearly a billion times higher than that of typical optical experiments~\cite{Aspect:1982fx,Weihs:1998gy,Hensen:2015ccp}, and within the domain of both strong and weak interactions.

Quantum entanglement and violations of the Bell inequality have also been recently confirmed in the charmonium sector, through the analysis of charmonium decays into baryon–antibaryon pairs--such as  $(\Lambda,\Sigma,\Lambda,\Xi)$ and  $(\phi)$ meson pairs~\cite{Fabbrichesi:2024rec}. These results were obtained using data on the helicity amplitudes of these processes from the BESIII Collaboration~\cite{BESIII:2018cnd,BESIII:2022qax,BESIII:2020fqg,BESIII:2024nif,BESIII:2021ypr,BESIII:2023drj,BESIII:2022lsz,BESIII:2023lkg,BESIII:2020lkm,BESIII:2023zcs,BESIII:2023euh}, as well as from the ATLAS~\cite{ATLAS:2014swk}, CMS~\cite{CMS:2018wjk}, and LHCb~\cite{LHCb:2013hzx} collaborations. These findings~\cite{Fabbrichesi:2024rec} provide another robust evidence for entanglement and quantum non-locality at the high energy scale of charmonium mass, in systems involving particles of different spins--in the qubit and qutrits bipartite systems--and governed by both electroweak and strong interactions.

In this chapter, the most significant results in particle physics concerning the measurement of quantum entanglement and the violation of Bell inequalities in bipartite qubit and qutrit systems are reviewed. Section~\ref{sec:tools} outlines the theoretical framework needed to analyze quantum observables in spin systems at particle colliders. Section~\ref{sec:entanglement-Bell} summarizes the main experimental results on entanglement and Bell inequality violations at colliders, with emphasis on studies performed at the LHC and at existing $e^+e^-$ machines, as well as on prospects for proposed future lepton colliders. Finally, Section~\ref{sec:conclusions} presents an outlook on future developments.

\section{Theoretical tools}
\label{sec:tools}
We present here the main quantum observables related to the concepts of Entanglement and Bell inequality pertaining to the measurement of spin correlations in bipartite quantum systems. We will focus here on bipartite systems made by pairs of  spin-1/2 (qubits) and spin-1 massive (qutrits) particles.
Central to the analysis is the formalism of the density matrix, $\rho$, which we introduce in the forthcoming section and apply to the polarization states of two qubits and qutrits particle systems.

\subsection{Density matrix formalism}
The statistical property of a quantum system is characterized by a set of quantum states $\ket{\psi_i}$--belonging to a Hilbert space--to which is associated a probability $p_i$ satisfying the conditions $0\le p_i\le 1$ and $\sum_i p_i=1$. In this case the state is said to be in a {\it mixed state}.
When the quantum state $\ket{\psi}$ is known exactly $(p=1)$, that is the set consists of a single state, the system is said to be in a  {\it pure state}.

The statistical description of mixed states, can be then described by an ensemble $\{(p_i,\ket{\psi_i})\}$ of pure states $\ket{\psi_i}$ and associated probabilities $p_i$ that encode our incomplete knowledge of the system. From this, we built an operator called \textit{density matrix}
\begin{equation}
  \rho = \sum_i p_i \op{\psi_i}\,,\qquad \sum_i p_i = 1\,,
\end{equation} 
which qualifies as a density operator  if and only if
\begin{enumerate}
  \item[i)] $\Tr[\rho]=1$
  \item[ii)] $\rho=\rho^\dagger$
  \item[iii)] ${\bra \psi} \rho {\ket \psi} \ge 0$~~~for all ~${\ket \psi}$\, ,
\label{rho-conditions}
\end{enumerate} 
where $\Tr[\rho]$ stands for the trace of $\rho$.
The normalization condition $i)$ represents the total probability (i.e., the system must be in some state with certainty), the  $ii)$ ensures that its eigenvalues are real, which is necessary since they represent probabilities, while the condition $iii)$ ensures that probabilities (expectation values) are non-negative.

The density matrix of a pure state is simply given by $\rho=\op{\psi}$, as we know exactly which of the states $\{\ket{\psi_i}\}$ actually describes the system. It then follows that for pure states
\begin{equation}
  \rho^2 = \rho\,,
\end{equation}
as the density matrix coincides with the $\ket{\psi}$ subspace projector.  Notice that in general $\Tr[\rho^2]\leq1$, with the equality holding only for pure states.

In the following we shall merely deal with bipartite composite quantum systems $S = S_A + S_B$ consisting of two finite-dimensional parties $S_A$ and $S_B$ , usually identified with two distant, well-separated quantum subsystems. An observable $O$ of the whole system $S$ can then be expressed in a tensor product form, $O= O_A \otimes O_B$ , where $O_A$ , $O_B$ are observables of $S_A$ and
$S_B$ , respectively.

Then, if the bipartite state $\rho$ of $S$ is separable, it can be expressed as a convex combination of two-qubit product states:
\be
\rho  = \sum_{ij} p_{ij}\,  \rho^{(A)}_i\otimes \rho^{(B)}_j\,,
\label{separable}
\ee
where $\rho^{(A)}_i$ and $\rho^{(B)}_j$ are the density matrices of the $A$ and $B$ systems separately, the coefficients $p_{ij}$ are strictly positive ($p_{ij}>0$) and they satisfy the normalization condition $\sum_{ij} p_{ij}=1$.
Any state whose density operator $\rho$ cannot be written in this form is referred to as {\it entangled}, and such states exhibit genuine quantum correlations.

\subsection{Entanglement and Bell non-locality}
\label{sec:entanglement-Bell-nonlocality}

Quantifying entanglement in composite quantum system is in general a difficult task~\cite{Horodecki:2009zz,gurvits2003classical,Gharibian:2008hgo}, especially in the case of mixed states, and only limited solutions are available.  There are however special cases where entanglement can be well quantified.
This is the case of pure states, where one can define a quantity called {\bf entropy of entanglement}~\cite{Horodecki:2009zz} $(E[\rho])$, also known as the von Neumann entropy of the reduced state, defined--for a bipartite system with subsystems $A$ and $B$ of the same dimensionality--as
\be
E[\rho] \equiv  -\Tr[\rho_A \ln\rho_A ] = -\Tr[\rho_B \ln \rho_B ]\, ,
\label{entropy}
\ee
where $\ln$ denotes the natural logarithm, and  the $\rho_A$ and $\rho_B$ are the reduced density matrices of subsystems $A$ and $B$, respectively. The entropy of entanglement is bounded as $0\le E[\rho]\le \ln d$, where $d$ is the finite dimension of the subsystems $A$ and $B$. The lower bound is saturated if and only if the bipartite pure state is separable (which, in the case of pure states, implies $\rho_{A,B}^2 = \rho_{A,B}$), while the upper bound is attained for a maximally entangled state. Clearly, a pure state is entangled if and only if its reduced density matrices have non-zero entropy.

In more general situations, as for instance in the case of mixed states, one must rely on so-called {\it entanglement witnesses}--quantities that provide sufficient conditions to indicate the presence of entanglement in the system. The most popular one is the so called \textbf{concurrence} ($\mathscr{C}$). In the case of a pure state it is  defined as~\cite{Bennett:PhysRevA.54.3824,Wootters:PhysRevLett.80.2245,Rungta:PhysRevA.64.042315}
\begin{equation}
\mathscr{C}[|\psi\rangle]\equiv\sqrt{2\left( 1-{\rm Tr}\big[(\rho_A)^2\big]\right)}
=\sqrt{2\left( 1-{\rm Tr}\big[(\rho_B)^2\big]\right)}\ ,
\label{C_psi}
\end{equation}
that vanishes if and only if the state is separable and it provides a true measure of entanglement.

Since any mixed state $\rho$ of a bipartite system can be decomposed  into a set of pure states $\{|\psi_i\rangle\}$, in the case of mixed states the concurrence is obtained as
\begin{equation}
\mathscr{C}[\rho]=\underset{\{|\psi\rangle\}}{\rm inf} \sum_i p_i\, \mathscr{C}[|\psi_i\rangle]\ ,
\label{C_rho}
\end{equation}
where the infimum (inf) has to be taken over all the possible decompositions of $\rho$ into pure states and it also vanishes for separable states.

Apart from few particular cases, such as the two-qubit systems--that will be discussed in the next session--the general solution to \eq{C_rho} is not known.
However, an analytical solution for the lower bound of the concurrence is known and it is given by~\cite{Mintert:PhysRevLett.98.140505}
\begin{equation}
\big(\mathscr{C}[\rho]\big)^{2} \geq \mathscr{C}_2[\rho]\ ,
\label{C-bound}
\end{equation}
where
\begin{equation}
\mathscr{C}_2[\rho] = 2 \,\text{max}\, \Big( 0,\, \Tr[\rho^{2}] - \Tr[(\rho_A)^2],\, \Tr[\rho^{2}] - \Tr[(\rho_B)^2]  \Big) \ .
\label{C_2}
\end{equation}
An upper bound for $\mathscr{C}[\rho]$ is also known~\cite{Zhang:PhysRevA.78.042308}
\begin{equation}
\big(\mathscr{C}[\rho]\big)^{2} \leq 2 \,\text{min}\, \Big(1 - \Tr[(\rho_A)^2],\  1 - \Tr[(\rho_B)^2] \Big)\, .
\label{C_2_upper}
\end{equation}
In the most general case of quantum correlations (of mixed states) at colliders, the lower bound (\ref{C_2}) on concurrence can be used as entanglement witness.

Entangled states are known to violate the so-called {\bf Bell inequalities}, which are mathematical expressions involving correlations between measurements performed on spatially separated parts of a bipartite quantum system. These inequalities serve as a crucial experimental tool for distinguishing between the predictions of quantum mechanics and those of classical theories based on local realism and determinism. A violation of a Bell inequality provides strong evidence that the behavior of entangled particles cannot be explained by any local hidden variable theory, thereby highlighting the fundamentally non-classical nature of quantum correlations.
The Bell inequalities are based on the concept of Bell locality which is satisfied by all deterministic theories.

Bell locality asserts that the measurement outcomes at one party (e.g., Alice) are independent of the choices and results at the other party (e.g., Bob), which are assumed to be far away separated without possibility to exchange causal information signals. Instead, all observed correlations arise from shared resources, denoted by a hidden variable $\lambda$, which may include the quantum wavefunction. Under this assumption, the joint probability $P(A,B|a,b)$ associated to the joint result $(a,b)$ of outcomes of $\hat A$ and $\hat B$ observables can be expressed as an integral over these shared resources:
\begin{equation}
P(A,B|a,b) = \int {\rm d}\lambda~ \eta(\lambda)\ P_\lambda(A|a)\ P_\lambda(B|b)\ ,
\label{Bell-locality}
\end{equation}
where $\eta(\lambda)$ is the probability distribution of the shared resources
and $P_{\lambda}(A|a)$ ($P_{\lambda}(B|b)$) the probability for Alice (Bob) of getting the outcome $a$ ($b$). 
This condition defines a {\it local} theory. If experimental results violate this expression--typically tested via Bell inequalities--the system is deemed {\bf non-local}, highlighting a fundamental departure from classical notions of locality.

A Bell test involves a set of inequalities that place constraints on the joint probabilities  $P(A,B|a,b)$ under the assumption of classical locality. A violation of these inequalities--consistently observed in experiments--demonstrates a clear departure from classical, local interpretations of nature, revealing the inherently non-local character of quantum phenomena.

While entanglement in pure states is always equivalent to Bell nonlocality, this equivalence does not generally extend to mixed states. In fact, some states are entangled yet still Bell local. Most examples of this kind trace back to the Werner state~\cite{Werner:1989zz}, which, in the case of a bipartite qubit system--such as a pair of spin-1/2 particles--is defined as a mixture of the singlet state and the identity. Another important example, discussed in Sec.~\ref{sec:top}, is the top–antitop quark system, which, like the Werner state, can be entangled but nevertheless Bell local~\cite{Fabbrichesi:2025psr}.

There are also other known correlations that can be used to probe quantum entanglement, such as quantum discord~\cite{Ollivier:2001fdq},
steerability~\cite{Wiseman:2007hyt}, and non-negative conditional entropy~\cite{Horodecki:2005vvo}. These observables capture different layers of quantum correlations beyond entanglement in the strict Bell sense. Quantum discord, for instance, identifies forms of non-classical correlation that remain even in separable states, while steerability characterizes the ability of one subsystem to non-locally influence the state of another through local measurements. Similarly, non-negative conditional entropy provides an information-theoretic criterion that reflects the presence of entanglement and its potential utility in quantum communication tasks. Together, these measures offer complementary perspectives on the richness of quantum correlations and extend the toolbox available for testing and quantifying entanglement in physical systems. A detailed discussion of these topics lies beyond the scope of this chapter, and we refer the reader to the literature for further details.

\subsection{Quantum observables with Qubits}
\label{sec:qubits}

In what follows, we identify spin systems with their corresponding quantum information analogues: a spin-$\tfrac{1}{2}$ particle corresponds to a qubit, a spin-1 particle to a qutrit, and more generally, a spin-$J$ particle to a qudit of dimension $2J+1$. The corresponding formalism requires the introduction of the polarization density matrix.

Let us begin with the definition of polarization density matrix of a single spin-$J$ particle. To this end, we introduce the polarized matrix element $M(\lambda)$ for the production of a spin-$J$ particle, where $\lambda$ generically denotes the spin eigenvalues along a chosen spin quantization axis—or, equivalently, the helicity eigenstates—with $\lambda = -J, -(J-1), \dots, (J-1), J$.
The corresponding polarization density matrix, a $(2J+1) \times (2J+1)$ matrix, is then defined as
\be
\rho(\lambda, \lambda^{\prime}) =
\frac{M(\lambda) M^{\dag}(\lambda^{\prime})}
     {|\overline{M}|^2} \, ,
     \label{density1}
\ee
where $|\overline{M}|^2$ denotes the corresponding unpolarized squared amplitude of the process and a sum over the possible internal degrees of freedoms of initial state particles is understood.

The generalization to two spin-$J$ particles is straightforward. In this case, we introduce the polarized matrix element $M(\lambda_1, \lambda_2)$ for the production of the two spin-$J$ particles, where $\lambda_1$ and $\lambda_2$ refer to the spin indices of particles 1 and 2, respectively. The corresponding polarization density matrix, a $(2J+1)^2 \times (2J+1)^2$ matrix, is then defined as
\be
\rho(\lambda_1, \lambda_1^{\prime}; \lambda_2, \lambda_2^{\prime}) =
\frac{M(\lambda_1, \lambda_2) M^{\dag}(\lambda_1^{\prime}, \lambda_2^{\prime})}
     {|\overline{M}|^2} \, ,
          \label{density2}
\ee
where again $|\overline{M}|^2$ denotes the unpolarized squared amplitude of the two spin-J particles. The technique to compute this polarization density matrix  for two spin-1/2 and spin-1 particles
will be provided in section \ref{poldens}.

Now, we start here with the discussion of the quantum observables related to the polarization density matrix for a pair  of two spin-half particles (qubits) denoted as $A$ and $B$. In this case, following the above considerations, see \eq{density2}, the quantum state of the corresponding bipartite system in the spin space can be modeled by the following Hermitian, normalized $4\times 4$ density matrix:

\be
\rho(\lambda_1, \lambda_1^{\prime}; \lambda_2, \lambda_2^{\prime})
= \frac{1}{4}\Big[ \mathbb{1} \otimes \mathbb{1} + \sum_i B_i^{\A} (\sigma_i\otimes \mathbb{1} )
+ \sum_j B_j^{\B}(\mathbb{1} \otimes \sigma_j)  
+ \sum_{ij} C_{ij} (\sigma_i\otimes\sigma_j) \Big]_{\lambda_1, \lambda_1^{\prime}; \lambda_2, \lambda_2^{\prime}}\ ,
\label{rhoqubit}
\ee
where $\sigma_i$ are Pauli matrices, $\lambda_{1,2}$ and $\lambda^{\prime}_{1,2}$ label the matrix elements, $\mathbb{1} $ denotes the $2\times 2$ identity matrix, and the sums of the indices $i$, $j$ run over the components of any orthonormal reference frame in three dimensions. The product
$A\otimes B$ generally indicates the usual Kronecker products among the $A$ and $B$ matrices. We use the abbreviation: $[A\otimes B]_{ii^\prime; jj^\prime} =
A_{ii^\prime} B_{jj^\prime}$.

The real coefficients $B_i^{\A}={\rm Tr}[\rho\, (\sigma_i\otimes \mathbb{1})]$ and $B_j^{\B}={\rm Tr}[\rho\, (\mathbb{1} \otimes\sigma_j)]$ represent the spin polarization of the two qubits (spin-1/2 particles $A$ and $B$, respectively), while the real matrix $C_{ij}={\rm Tr}[\rho\, (\sigma_i\otimes\sigma_j)]$ gives their spin-correlations. 
In the case of the particle pair system, $B_i^{\A,\B}$ and $C_{ij}$ are functions of the parameters describing the kinematics of the pair production. In addition, these coefficients need to satisfy further constraints coming from the positivity request, $ii)$ condition in \eq{rho-conditions}, that any density matrix should fulfill; these extra conditions are in general non-trivial, as they originate from requiring all principal minors of the matrix $\rho$ to be non-negative.

In the special case of two qubits the entanglement can be quantified for any quantum state. Indeed, the concurrence $\mathscr{C}[\rho]$  can be analytically evaluated by using the auxiliary matrix
\begin{equation}
R=\rho \,  (\sigma_y \otimes \sigma_y) \, \rho^* \, (\sigma_y \otimes \sigma_y)\, , 
\label{auxiliary-R}
\end{equation}
where $\rho^*$ denotes the matrix with complex conjugated entries. Although non-Hermitian, the matrix $R$ has non-negative eigenvalues; denoting with $r_i$, $i=1,2,3,4$, the square roots of them and assuming $r_1$ to be the largest,
the concurrence of the state $\rho$ is given by~\cite{Wootters:PhysRevLett.80.2245}
\begin{equation}
  \mathscr{C}[\rho] = \max \big( 0, r_1-r_2-r_3-r_4 \big)\ .
\label{concurrence}
\end{equation}

Now we present the main observable adopted to test the
{\bf Bell non-locality} in qubit systems. In this case one can use of the following correlator for the pair of observables $(a_1, a_2; b_1, b_2)$ related to the expectation values of the operators $(\hat{A}_1,\hat{A}_2)$ and $(\hat{B}_1,\hat{B}_2)$ by Alice and Bob observers respectively~\cite{Clauser:1969ny}:
\begin{equation}
\mathcal{I}_2=\langle {\hat A}_1 {\hat B}_1\rangle + \langle {\hat A}_1 {\hat B}_2\rangle +\langle {\hat A}_2 {\hat B}_1\rangle
- \langle {\hat A}_2 {\hat B}_2\rangle\, ,
\label{CHSH}
\end{equation}
where $\langle {\hat A}_i {\hat B}_j\rangle$ stand for the expectation value of the observable ${\hat A}_i {\hat B}_j$ and 
with each observables giving two possible outcome (0, 1) of these measurements. Then, the following  {\bf Clauser-Horne- Shimony-Holt (CSHS) inequality}
holds~\cite{Clauser:1969ny}:
\begin{equation}
\mathcal{I}_2\leq 2\ .
\label{CHSH-2}
\end{equation}
A violation of the \eq{CHSH-2} inequality signals the presence of Bell non-locality that can rule out any deterministic local expectations.

Now, we associate the operators $\hat{A}_{1,2}$ and $\hat{B}_{1,2}$ to the spin projections along four different unit vector directions, say $\vec{n}_1$, $\vec{n}_3$ for Alice, and $\vec{n}_2$, $\vec{n}_4$ for Bob, in such a way that 
$\hat{A}_1= \vec{n}_1\cdot \vec{\sigma}$ etc. 
Then, in this case, the CHSH-2 inequality above \eq{CHSH-2} can be compactly rewritten by using the density matrix $\rho$ as
\be
\mathcal{I}_2={\rm Tr}[\rho \mathscr{B}]
\ee
where the quantum \textbf{Bell operator} $\mathscr{B}$ is given by
$\mathscr{B} ={\vec n}_1 \cdot {\vec\sigma} \otimes ({\vec n}_2 - {\vec n}_4) \cdot {\vec\sigma}
+ {\vec n}_3 \cdot {\vec\sigma} \otimes ({\vec n}_2 + {\vec n}_4 )\cdot  {\vec\sigma}\ .
$

The Bell inequality (\ref{CHSH-2}) can be then expressed as a function of the correlation matrix $C$ in \eq{rhoqubit} as 
\begin{equation}
{\vec n}_1\cdot C \cdot \big({\vec n}_2 - {\vec n}_4 \big) +
{\vec n}_3\cdot C \cdot \big({\vec n}_2 + {\vec n}_4 \big)\leq 2\ .
\label{algebraic-relation}
\end{equation}
The optimization problem related to the possible choice of four ${\vec n}_i$ directions that maximizes \eq{algebraic-relation} above, can be performed for the most general correlation $C$ matrix. In particular, assuming $m_1$, $m_2$ are the largest eigenvalues of the three eigenvalues ($m_1$, $m_2$, $m_3$) of the matrix $M= C C^T$, and if
the sum of the two greatest eigenvalues of $M$ matrix satisfies the following condition (\textbf{Horodecki condition}) \cite{Horodecki:1995nsk}
\begin{equation}
\mathfrak{m}_{12}\equiv m_1 + m_2 >1\, ,
\label{eigenvalue-inequality}
\end{equation}
then there exists directions ${\vec n}_i$ directions that maximizes the violation of the Bell inequality in \eq{algebraic-relation} above.
Notice, the sufficient condition above should be strictly larger than 1.

 In the following phenomenological studies, we will refer to the $\mathfrak{m}_{12}>1$ condition above for the Bell inequality violations in the two qubit systems.

\subsection{Quantum observables with Qutrits}
\label{sec:qutrits}

Here we analyze the case of the bipartite state made by  the polarization density matrix for a pair of two spin-1 particles which can be associated to two qutrits. Following the considerations in section \ref{sec:qubits}, see \eq{density2}, the polarization density matrix describing the joint state of two qutrits is then a $9\times 9$ matrix. As such, it is convenient to decompose the corresponding density matrix as a sum of tensor products involving the Gell-Mann matrices, $T^a$ with $a=1,\dots,8$ and the $3\times3$ unit matrix $\mathbb{1}_3$ as follows
\bea
\label{eq:rhone}
\rho
= \Big[
  \frac{\mathbb{1}_3\otimes
  \mathbb{1}_3}{9}
    +
    \sum_{a=1}^8 f_a (T^a \otimes \mathbb{1}_3)
    +
    \sum_{a=1}^8 g_a (\mathbb{1}_3\otimes T^a) 
    +\sum_{a,b=1}^8 h_{ab}  (T^a\otimes T^b)\Big]\, ,
\label{rho-qutrit}
\eea
where we omitted the spin indices, and $A\otimes B$ stands for the usual Kronecker products as defined above. The coefficients $f_a$ and $g_a$, $a=1,\dots,8$, as well as the 64 elements of the symmetric matrix $h_{ab}$, generally depend on the kinematics of the process yielding the production of the two qutrits and can be computed analytically from the related transition amplitudes. 

These coefficients can be obtained by projecting $\rho$ in \eq{rho-qutrit} on the desired subspace basis via the traces
\be
f_a=\frac{1}{6}\,\Tr\left[\rho \left(T^a \otimes \mathbb{1}_3\right)\right]\, , ~~
g_a=\frac{1}{6}\,\Tr\left[\rho \left(\mathbb{1}_3\otimes T^a\right)\right]\, , ~~h_{ab}=\frac{1}{4}\,\Tr\left[\rho \left(T^a \otimes T^b\right)\right]\,.
\label{fgh}
\ee
The density matrix in \eq{rho-qutrit} can be also expanded in the basis formed by the tensor products of irreducible tensor operators \cite{Aguilar-Saavedra:2022wam}.

An alternative way to express the density matrix is through its decomposition in terms of tensor operator components, which is given by~\cite{Aguilar-Saavedra:2015yza,Bernal:2023jba}
\be
\rho = \frac{1}{9}\Bigg\{ \mathbb{1}\otimes
    \mathbb{1}+
    A^1_{LM} \left[T^L_M\otimes \mathbb{1}\right]+ A^2_{LM}  \left[\mathbb{1}\otimes T^L_M \right] 
    +C_{L_1 M_1 L_2 M_2}  \left[T^{L_1}_{M_1}\otimes T^{L_2}_{M_2}\right]\Bigg\}\, .
\label{eq:rho-tensor2}
\ee
where $T_M^L$ are the matrices corresponding to irreducible spherical tensor operators. By means of the following orthogonality relation 
$\Tr{\left[T_{M'}^{L'} T_M^{L\dag}\right]} = \left(\frac{2s+1}{2L+1}\right) \delta_{LL'} \delta_{MM'}$ the coefficients $A^1_{LM} = \Tr{\left[\rho~ T_M^L \otimes \mathbb{1} \right]}$, $A^2_{LM} = \Tr{\left[\rho~\mathbb{1} \otimes  T_M^L \right]}$,
$C_{L_1 M_1 L_2 M_2} = \Tr{\left[\rho~T^{L_1}_{M_1}\otimes T^{L_2}_{M_2}\right]}$ can be extracted from the polarization density matrix.

In the case of two qutrits quantifying entanglement for any quantum state is a very challenging task and no analytical solutions for the concurrence $\mathscr{C}[\rho]$  are known for the general case.  However, in  this case one can always rely on the entanglement witness provided by the lower bound $\mathscr{C}_2$ of the concurrence, as given in \eq{C_2}.
This can be written in terms of the $f_a,g_a,h_{ab}$ coefficients in \eq{fgh} as
\bea
\mathscr{C}_2
&=& 2\max \Big[ -\frac{2}{9}-12 \sum_a f_{a}^{2} +6 \sum_a g_{a}^{2} + 4 \sum_{ab} h_{ab}^{2}
  -\frac{2}{9}-12  \sum_a g_{a}^{2} +6 \sum_a f_{a}^{2} + 4 \sum_{ab} h_{ab}^{2},\, 0 \Big]\, ,
~~~~\label{C2}
\eea
which is the expression we use throughout this work.

In the special case of pure states, one can always quantify entanglement of two qutrits by using the entropy of entanglement, as given in \eq{entropy}. In the following, we will provide an analitycal expression of the entropy in terms of the polarization coefficients only in the special cases of a scalar decay into two spin1.

\vspace{0.5cm}
{\bf Bell locality test for qutrits}: a reformulation of the original Bell inequality, tailored specifically for a bipartite system composed of two qutrits is provided by  the {\bf Collins, Gisin, Linden, Massar, and Popescu (CGLMP) inequality} ~\cite{Collins:2002sun,Kaszlikowski:PhysRevA.65.032118, Horodecki:1995340,Brunner:RevModPhys.86.419}. To express this inequality explicitly, consider a system consisting of two qutrits, labeled $A$ and $B$. For qutrit $A$, choose two spin measurement settings, denoted by $\hat{A}_1$ and $\hat{A}_2$ , each corresponding to projective measurements of spin-1 observables with three possible outcomes: $\{0, 1, 2\}$. Similarly, for qutrit $B$, the measurement settings are represented by
$\hat{B}_1$ and $\hat{B}_2$. Based on these observables, one can define a correlation function, $I_3$, constructed through a specific combination introduced in~\cite{Collins:2002sun}:
  \bea
\II &=& P(A_1 = B_1 ) + P(B_1 = A_2 + 1) + P(A_2 = B_2) + P(B_2 = A_1)
\nonumber \\
&-&P(A_1 = B_1 - 1) - P(A_1 = B_2)- P(A_2 = B_2 - 1) - P(B_2 = A_1 - 1) \,,
\label{I3}
\eea
in which $P (A_i = B_j + k)$  stands for the probability that the outcome $A_i$ (for the measurement of $\hat{A}_i$) and $B_j$ (for the measurement of $\hat{B}_j$), with $i, j$ either 1 or 2, differ by $k$ modulo 3. Deterministic local models inevitably yield
\be
\II \le 2\, ,
\ee
while the bound can be violated by computing the above correlation measure with the rules of quantum mechanics.

Within the density matrix framework, the quantity above can be expressed as the expectation value of a suitable Bell operator $\Bell$ for qutrits
\cite{Collins:2002sun}
\be
\II = \Tr[\rho \Bell]\, ,
\label{I3}
\ee
where the precise form of $\Bell$ depends on the choice of the four measurement operators $\hat{A}_i$, $\hat{B}_i$, with $i = 1,2$, used in the test. For the case of the maximally correlated qutrit state,
the problem of finding an optimal choice of measurements has been solved~\cite{Collins:PhysRevLett.88.040404}
and the $9\times 9$ matrix of the Bell operator $\Bell$ assumes a simple form, its expression can be found in~\cite{Latorre:PhysRevA.65.052325,Barr:2024djo}.

Therefore, for a given density matrix, it is possible to increase the violation of the Bell inequality by an appropriate selection of these operators. However, irrespective of this choice, the resulting value of the observable must not exceed $4$ if quantum mechanics is to remain valid.

Moreover, the measurement settings that define the Bell operator can be further adjusted through local unitary transformations of the observables. Under such transformations, the Bell operator changes according to:
\be
\Bell \to (U \otimes V)^{\dag} \cdot \Bell \cdot (U \otimes V)\ , \label{uni_rot}
\ee
where $U$ and $V$ are independent unitary matrices of dimension $3 \times 3$. This process effectively represents local basis rotations applied to the measurements on each qutrit. In what follows, the phenomenological analysis presented in the next section \ref{sec:entanglement-Bell} exploits this freedom to optimize the value of ${\cal I}_3$ obtained from the Bell operator as defined in Ref.~\cite{Acin:2002zz}, in order to maximize the Bell inequality violation for $\II$.

\subsection{Polarization Density matrix}
\label{poldens}

In this section, we present the Lorentz covariant method for computing the polarization density matrix for spin-half fermions and massive spin-one particles--these are the cases that will serve as the basis for our phenomenological analysis in what follows. A separate discussion based on the helicity amplitude method will follow thereafter.
\subsubsection{Spin-1/2 polarization matrix}
\label{sec:rho-spin-half}
Here we provide the tool box for computing the density matrix related to the polarization state $\lambda$ of a spin-half fermion  $\psi_\lambda$. Generalization to a bipartite state of two spin-half fermion is straightforward.

We start with the polarized amplitude $\mathcal{M}(\lambda)$ of for production process of a spin-half fermion  $\psi_\lambda$
	\begin{equation}
		\mathcal{M}(\lambda) = \qty[\bar u_\lambda \mathcal{A}],
	\end{equation}
with $\lambda$ indicating the polarization $\lambda\in\{-\frac{1}{2},\frac{1}{2}\}$ along a specified quantization direction, where $\mathcal{A}$ the term in the amplitude that multiplies the spinor $\bar u_\lambda$ of the produced fermion. We adopt the notation of square brackets $[\cdots ]$ to track the contractions of spinor indices for the quantities inside them.

Then, the density matrix is given by \cite{Barr:2024djo}
	\begin{equation}\label{project-spin-half}
		\rho_{\lambda \lambdap}=\frac{\qty[\mathcal{A}\mathcal{A}^\dagger \Pi^u_{\lambda\lambdap}]}
		{\abs{\mathcal{M}}^2}
		\equiv\frac{1}{2}\left(\mathbb{1}+\sum_{i=1}^3 s^i\sigma_i\right)_{\lambda \lambdap}\, ,
	\end{equation} 
where $s^i$ are the components of the fermion polarization vector that generally depend on the kinematic variables of the underlying production process.
Above, the $\Pi^u_{\lambda\lambdap}$ ($\Pi^v_{\lambda\lambdap}$) quantities are the projection operators for fermion (anti-fermion) particle ~\cite{Bouchiat:1958yui,Leader2001}
	\be
		\label{eq:projU}
		\Pi^u_{\lambda\lambdap}= \frac{1}{2}\qty(\slashed{p}+m)\qty(\delta_{\lambda \lambdap}+\gamma_5\sum_i\slashed{n}_i\sigma^i_{\lambdap \lambda})
 \, , ~~~\Pi^v_{\lambda\lambdap}= \frac{1}{2}\qty(\slashed{p}-m)\qty(\delta_{\lambda \lambdap}+\gamma_5\sum_i\slashed{n}_i\sigma^i_{\lambda \lambdap})\, ,
	\ee
        where $\slashed{p}\equiv \gamma_{\mu} p^{\mu}$--with $\gamma_{\mu}$ and
$\gamma_5$  the usual Dirac matrices--the $\sigma_i$ are the Pauli matrices and $\{n_i^\mu\}$ is a triad of space-like four-vectors, each satisfying $n_i^\mu p_\mu =0$. The $n_i^\mu$ are obtained by boosting the canonical basis of the spin four-vector $n=(0, \vec n)$
        in the rest frame of the fermion to the frame where the fermion has four-momentum $p$. Above $\vec n$  defines the spin polarization vector (of norm 1) in the rest frame of the fermion.
        
        The generalization to processes involving multiple spin-half fermions in the final state is straightforward. The resulting density matrices can be expanded using a basis formed by tensor products of Pauli matrices and the identity matrix. In particular, in this case if we define 
\begin{equation}
		\mathcal{M}(\lambda) = \qty[\bar u_{\lambda_2} \mathcal{\hat{A}}~u_{\lambda_1}],
	\end{equation}
then we have 
	\begin{equation}\label{project-2-spin-half}
		\rho_{\lambda_1 \lambdap_1,\lambda_2 \lambdap_2}=\frac{\qty[\mathcal{\hat{A}}\mathcal{\hat{A}}^\dagger \Pi^u_{\lambda\lambdap}\otimes  \Pi^u_{\lambda\lambdap}]}
		    {\abs{\mathcal{M}}^2}\, ,
	\end{equation} 
Here, the tensor product is implicitly applied to the Pauli matrices associated with the two fermions, with spin index contractions understood throughout.
This leads to the bipartite density matrix given in \eq{rhoqubit}, with its components $B^{\A,\B}_i$ and $C_{ij}$ that can be extracted by computing the trace of the $\rho$ in \eq{project-2-spin-half} with the corresponding projection operator of the polarization and correlation matrix, as shown below \eq{rhoqubit}.

The following section \ref{sec:entanglement-Bell}, focused on the quantum tomography of top-quark and $\tau$-lepton pair production, we will provide further details on the derivation of polarization and correlation coefficients within the specific context of collider experiments.

\subsubsection{Spin-1 polarization matrix}
\label{sec:rho-spin1}
Here we consider spin-1 massive particles whose polarizations identify systems of qutrits.
In analogy with the spin-half system described above, 
we decompose the amplitude for the production of a massive gauge boson with helicity $\lambda\in\{+1, 0, -1\}$ and momentum $p$ as
\begin{equation}
    \mathcal{M}(\lambda, p) = \mathcal{A}_\mu \varepsilon^{\mu*}_\lambda(p)
    \label{eq:amp} 
\end{equation}
where $\varepsilon^{\mu\star}_\lambda$ denotes the (conjugated) polarization vector  of the produced boson.   
From the orthonormality relation $g_{\mu\nu}\,\varepsilon^\mu_\lambda(p) \,\varepsilon^\nu_{\lambda'}(p)=-\delta_{\lambda\lambda'}$ and Eqs.~\eqref{eq:amp} it follows that
\begin{equation}
    \rho_{\lambda\lambda'}=\frac{\mathcal{M}(\lambda)\mathcal{M}^\dagger(\lambda')}{\sum_{\lambda''} \mathcal{M}^\dagger(\lambda'')\mathcal{M}(\lambda'')}
    =\frac{\mathcal{A}_\mu \mathcal{A}^\dagger_\nu \mathcal{P}^{\mu\nu}_{\lambda\lambda'}}{\abs{\mathcal{M}}^2}\equiv\frac{1}{3}\left(\mathbb{1}+\sum_{i=1}^3 f_aT^a\right)_{\lambda \lambdap}\, ,
    \label{eq:rpol2}
\end{equation}
where $\mathcal{P}^{\mu\nu}_{\lambda\lambda'}(p)=\varepsilon^{\mu*}_\lambda(p) \, \varepsilon^{\nu}_{\lambda'}(p) \, $ is the projector and $\abs{\mathcal{M}}^2$ stands for the unpolarized contribution.
Above, the $f_a$ are the components of the spin-1 polarization vector that generally depend on the kinematic variables of the underlying production process.

In order to obtain an expression for the projector $\mathcal{P}$, consider the explicit form of the wave vector of a massive gauge boson with helicity $\lambda$
\begin{equation}
    \varepsilon^{\mu}_\lambda (p)=-\frac{1}{\sqrt{2}}|\lambda|\left(\lambda \, n_1^{\mu}+i \, n_2^{\mu}\right)
+\Big(1-|\lambda| \Big)n_3^{\mu} ,
\label{eps}
\end{equation}
where the four-vectors $n_i=n_i(p)$, $i\in\{1,2,3\}$, form a right-handed triad and are obtained by boosting the linear polarization vectors defined in the frame where the boson is at rest to a frame where it has momentum $p$, which in example above is chosen to be along the $n_3$ direction. With the above expression one finds~\cite{Kim1980,Choi1989,Fabbrichesi:2023cev}
\begin{equation}
\mathcal{P}^{\mu\nu}_{\lambda \lambda '}(p) =
\frac{1}{3}\left(-g^{\mu\nu}+\frac{p^{\mu}p^{\nu}}{m_V^2}\right)
\delta_{\lambda\lambda '}-\frac{i}{2m_V}
\epsilon^{\mu\nu\alpha\beta}p_{\alpha} n_{i\,\beta} \left(S_i\right)_{\lambda\lambda '}-\frac{1}{2}n_i^{\mu}n_j^{\nu} \left(S_{ij}\right)_{\lambda\lambda '},
\label{proj}
\end{equation}
where $m_V$ is the invariant mass of the vector boson $V$, $\epsilon^{\mu\nu\alpha\beta}$ the permutation symbol ($\epsilon^{0123}=1$) and $S_i$, $i\in\{1,2,3\}$, are the $SU(2)$ generators in the spin-1 representation--the eigenvectors of $S_3$, corresponding to the eigenvalues $\lambda\in\{+1, 0, -1\}$, define the helicity basis. The spin matrix combinations appearing in the last term are given by
\begin{equation}
S_{ij}= S_iS_j+S_jS_i-\frac{4}{3} \mathbb{1}\, \delta_{ij},
\label{Sij}
\end{equation}
with $i,j\in\{1,2,3\}$ and $\mathbb{1}$ being the $3\times 3$ unit matrix. Then, the 
$S_i$ and $S_{ij}$ matrices can be easily decomposed into the Gell-Mann basis $T^a$, see for instance \cite{Barr:2024djo}.

The formalism can be readily generalized to processes that produce a bipartite qutrit state composed of two massive gauge bosons, $V_1$ and $V_2$. In this case, we obtain
\begin{equation}
    \rho=\frac{{\mathcal A}_{\mu\nu}{\mathcal A}^{\dagger}_{\mu'\nu'}}{\abs{\mathcal{M}}^2}
    \left( \mathcal{P}^{\mu\mu'}(k_1)\otimes\mathcal{P}^{\nu\nu'}(k_2) \right),
    \label{eq:rhoVV}
\end{equation}
where $k_1$ and $k_2$ denote the momenta of the vector bosons in a given frame.  and the tensorial product $\otimes$ apply to the corresponding $T^a$ and unit matrix associated to  the two spin-1 particles.

The eight components of $f_a$ and $g_a$, along with the 64 elements of $h_{ab}$, can be extracted by projecting the density matrix in \eq{rho-qutrit}   onto the appropriate basis of the subspace, using orthogonality relations. This yields
\bea
f_a=\frac{1}{6}\,\Tr\Big[\rho \left(T^a \otimes \mathbb{1}\right)\Big]\, , ~~
g_a=\frac{1}{6}\,\Tr\Big[\rho \left(\mathbb{1}\otimes T^a\right)\Big]\, , ~~
h_{ab}=\frac{1}{4}\,\Tr\Big[\rho \left(T^a \otimes T^b\right)\Big]\, .
\label{fgh}
\eea
All the terms computed via Eq.~(\ref{fgh}) are Lorentz scalars.

We stress here that the Gell-Mann representation of the density matrix in
\eq{rho-qutrit} is only one possible parametrization. For alternative representations of the density matrix of qutrits in terms of tensor operator components see \cite{Aguilar-Saavedra:2015yza,Bernal:2023jba}.


\subsubsection{Helicity-amplitude spin formalism}
\label{sec:helicity-amplitude}

We now discuss the connection with the helicity amplitude spin formalism. 
\textit{Helicity amplitudes} are defined as the matrix elements of the $S$-matrix taken between initial and final helicity eigenstates. These helicity states are typically specified in the center-of-mass frame, where the scattering angle determines the quantization axis used to define the helicity eigenstates $\ket{\lambda_i}$ of the final-state particles. These states reduce to the usual spin eigenstates when boosted to the rest frame of the $i$-th particle.

By decomposing the $S$-matrix as $S = 1 + iT$, the transition amplitude can be written as
\begin{equation}
  \delta^{(4)}(p_1 + p_2 - k_1 - k_2)\, {\cal M}(\lambda_A, \lambda_B) \propto \mel{\Omega(\theta,\phi), \lambda_A\, \lambda_B}{T}{\Omega(0,0), \chi_1\, \chi_2} \,,
\end{equation}
$k_1$ and $k_2$ are the momenta of the initial-state particles, which are taken along the $z$-axis, i.e., $\Omega(\theta=0,\phi=0)$, with $\theta$ and $\phi$ denoting the polar and azimuthal angles, respectively, and  the whole dependence on scattering angles is absorbed in the $\Omega(\theta,\phi)$ term.
Above, $\chi_1$ and $\chi_2$ denote the helicities of the initial-state particles, while, as before, $\lambda_A$ and $\lambda_B$ refer to the helicities of the final-state particles, in the center of mass frame. 

The transition amplitude can then be expanded in partial waves as
\begin{equation}
  \mel{\Omega(\theta,\phi), \lambda_A\, \lambda_B}{T}{\Omega(0,0), \chi_1\, \chi_2}
  = \frac{1}{4\pi} \sum_J (2J+1) \mel{\lambda_A \lambda_B}{T^J}{\chi_1 \chi_2} \, \mathcal{D}^{J*}_{\chi \lambda}(\phi, \theta, 0) \,.
\end{equation}
As we can see, the whole angular dependence of the amplitude is contained in the $\mathcal{D}^J_{\chi,\lambda}(\alpha, \beta, \gamma)$ terms which are the Wigner $D$-matrix elements in the spin-$J$ representation of the rotation group
(with $\alpha, \beta, \gamma$ the usual Euler's angles), with $\chi = \chi_1 - \chi_2$ and $\lambda = \lambda_A - \lambda_B$. 

The helicity density matrix for the final state can then be written as
\begin{align}
  \label{eq:rhoHA}
  \rho_{\lambda_A \lambda_B; \lambda_A' \lambda_B'} 
  &=
  \frac{1}{\abs{\mathcal{M}}^2} 
  \sum_J w_{\lambda_A \lambda_B}^J \, w^{*J}_{\lambda_A' \lambda_B'} \sum_{k=-J}^J  
  \mathcal{D}^{J*}_{k, \lambda_A - \lambda_B}(0, \theta, 0) \, 
  \mathcal{D}^J_{k, \lambda_A' - \lambda_B'}(0, \theta, 0) \,,
\end{align}
where a sum over the helicities of the initial-state particles is implicitly understood, and $w_{\lambda_A \lambda_B}^J \propto \mel{\lambda_A \lambda_B}{T^J}{\chi_1 \chi_2}$. The overall normalization factor is fixed by the condition $\operatorname{Tr}(\rho) = 1$, and the sum over all polarizations in the square amplitude of the process at fixed helicities ($\abs{\mathcal{M}}^2$) is understood. As expected from the cylindrical symmetry of the process, the dependence on the azimuthal angle $\phi$ cancels out.

A similar expression holds for $1 \to 2$ decay processes, in which case $J$ corresponds to the spin of the decaying particle. As we will see in the following, the expression in \eq{eq:rhoHA} proves particularly useful for reconstructing the density matrix in this case.

\subsection{Reconstructing the polarization density matrix from decays}
The polarization density matrix of a spin-particle system can be experimentally reconstructed from the angular distributions of its decay products, provided these distributions are accessible in the rest frame of the parent particles.
Here, we focus on two specific cases: the production of a pair of spin-$\frac{1}{2}$ particles (qubits) and a pair of spin-1 particles (qutrits), assuming that the parent particles decay into detectable daughter products.

\subsubsection{Qubits}
\label{sec:simulation-qubits}
As an example, consider the production of a top-quark pair, $t\bar{t}$, and their decay into the leptonic channel:
$t\bar{t} \to \ell^+ \nu b ~ \ell^- \bar{\nu} \bar{b}$,
where $\ell^{-}(\ell^{+})$ denotes the lepton (antilepton), $\nu (\bar{\nu})$ the neutrino (antineutrino), and $b (\bar{b})$ the bottom (antibottom) quark.
This method can be generalized to any other fermion-antifermion pair.

We adopt the same normalized right-handed basis $\{{\bf {\hat{r}, \hat{n}, \hat{k}}}\}$ in both the top and antitop rest frames as proposed in \cite{Bernreuther:2015yna,Bernreuther:2010ny}. Here, ${\bf \hat{k}}$ represents the direction of the top-quark motion in the $t\bar{t}$ center-of-mass frame, and ${\bf \hat{p}}$ denotes the direction of one of the proton beams in the laboratory frame. The remaining directions, ${\bf \hat{n}}$ and ${\bf \hat{r}}$, are then defined as follows\cite{Bernreuther:2015yna,Bernreuther:2010ny}:
\bea
    {\bf \hat{n}}&=&\frac{1}{\sin{\Theta}} {\bf \hat{p}} \times {\bf \hat{k}}
    \nonumber \\
              {\bf \hat{r}}&=&\frac{1}{\sin{\Theta}}
              \left({\bf \hat{p}}- {\bf \hat{k}}\cos{\Theta}\right)\, ,
\label{basis-rnk}
              \eea
where $\cos{\Theta}\equiv {\bf \hat{p}}\cdot {\bf \hat{k}}$ is the angle between the ${\bf \hat{p}}$ and ${\bf \hat{k}}$ directions, and ${\bf \hat{n}}$ is the direction perpendicular to their plane. We then define the leptonic angles
$\theta^i_{\pm}$ as
\bea
\cos{\theta^i_{\pm}}=\hat{p}_{\ell^{\pm}}\cdot  {\bf \hat{\it i}}\, ,
\eea
where ${\bf \hat{\it i}}=\{\bf \hat{r},\hat{n},\hat{k}\}$, and
$\hat{p}_{\ell^{+}}$ ($\hat{p}_{\ell^{-}}$) are the momenta of charged leptons
$\ell^+ (\ell^-)$ in the corresponding top (antitop) rest frames.
See Fig.\ref{fig:frame} for the production of a generic fermion pair $\psi (\bar{\psi})$ decaying into leptons $\ell^- (\ell^+)$. illustration.

\begin{figure}[h!]
\begin{center}
\includegraphics[width=3.in]{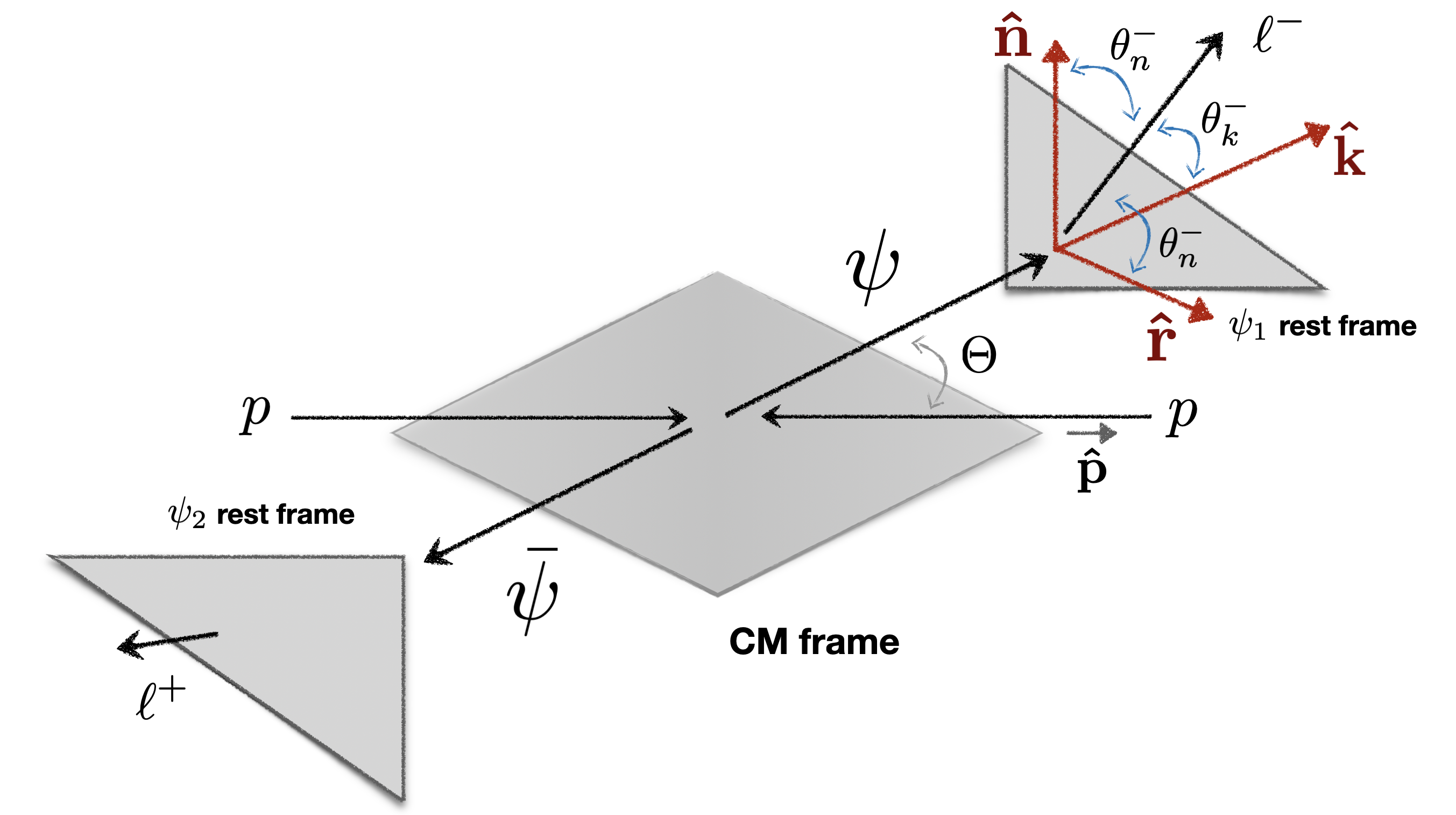}
\caption{\small Unit vectors and momenta in the CM system~\cite{Bernreuther:2001rq}, here specified for the production $p\, p \to \psi \bar \psi$. The angles $\theta^-_i$ define the directions of the final lepton in the rest frame of the fermion $\psi$ with respect to the quantization axis. The same holds for $\bar \psi$, while $\Theta$ is the scattering angle in the $\psi-\bar{\psi}$ center of mass frame. (Figure from \cite{Barr:2024djo} under the \href{https://creativecommons.org/licenses/by/4.0/}{CC BY 4.0} licence).
\label{fig:frame} 
}
\end{center}
\end{figure}

In terms of these definitions, the normalised differential distribution is given by \cite{Bernreuther:2015yna}
\bea
\frac{1}{\sigma} \frac{d \sigma}{d \cos{\theta_{+}^i}
      d \cos{\theta_{-}^j}}&=& \frac{1}{4}\left( 1+\kappa_+ B_i^{+} \cos{\theta_{+}^i}
+\kappa_- B_j^{-} \cos{\theta_{-}^j}+\kappa_+ \kappa_- C_{ij}\cos{\theta_{+}^i}\cos{\theta_{-}^j}\right)\, ,
\label{distr:ttbar}
\eea
where ${\bf \hat{\it j}}=\{\bf \hat{r},\hat{n},\hat{k}\}$, and the $C_{ij}$ and $B_i^{\pm}$ coefficients are the same coefficients appearing in the density matrix of two qubits in \eq{rhoqubit}. The coefficients $\kappa_{\pm}$ 
are the spin analyzing powers--which quantify the amount of transferred polarization from the parent to the daughter decay particle--that for the top quark pairs are $\kappa_+=+1$ and $\kappa_=-1$ for the positive and  negative leptons respectively.

In general, the spin analyzing powers can be measured from the angular distribution of the decay product $f$ of a polarized parent particle, in the rest frame of the parent particle:
\bea
\frac{1}{\Gamma} \frac{d \Gamma}{d \cos{\theta}}&=& \frac{1}{2}\left(1+\kappa_f\ P \cos{\theta}\right)\, ,
\eea
where $\Gamma$ is the width, $\theta$ is the angle between the spin direction of the parent particle and the momentum of the decay product $f$, $\kappa_f$ is the spin analyzing power of the decay product  that ranges from -1 to 1, and $P$ is the polarization of the parent particle.

Then, by fitting the data with double distribution $\cos{\theta_{+}^i} \cos{\theta_{-}^j}$ in \eq{distr:ttbar} along the $ij$ directions, one can extract the correlation coefficients $C_{ij}$. Similarly, fitting the single distributions $\cos{\theta_{+}^i}$ and $\cos{\theta_{-}^j}$ allows the determination of the polarization coefficients $B^i_{+}$ and $B^j_{-}$, respectively.

\subsubsection{Qutrits}
\label{sec:simulation-qutrits}
We treat spin-1 gauge bosons as qutrits, with their decay products serving as built-in polarimeters. We start as an example with the positively charged weak boson $W^{+}$. The polarization of the $W^+$ can be reconstructed via the decay $W^+ \to \ell^+ \nu$. In this case, the lepton $\ell^+$ is emitted with positive helicity and the neutrino with negative helicity, indicating a $W^+$ polarization of +1 along the lepton direction. Similarly, in $W^- \to \ell^- \bar{\nu}$, the lepton $\ell^-$ has negative helicity and the antineutrino positive, yielding a $W^-$ polarization of -1 along the lepton direction.

In both cases, the direction of the final-state leptons (as for instance in
Fig.\ref{fig:frame}) provides a direct probe of the W boson polarization. These particle momenta are the essential observables required--either from simulation or experimental data--to determine the polarization states of the gauge bosons.

Using the method outlined in~\cite{Ashby-Pickering:2022umy}, the correlation ($h_{ab}$) and polarization ($f_a, g_a$) coefficients from the density matrix decomposition in \eq{rho-qutrit} can be extracted via the double differential angular distribution~\cite{Rahaman:2021fcz}:
\be
\frac{1}{\sigma}\frac{\di \sigma}{\di\Omega^{+},\di\Omega^{-}}
= \left( \frac{3}{4 \pi} \right)^2
\Tr \Big[ \rho_{V_1V_2} \left(\Pi_+ \otimes \Pi_-\right)\Big]\, ,
\label{x-sec-leptons}
\ee
where $\sigma$ is the production cross section of the two gauge bosons decaying into the leptonic modes, and $\di \Omega^\pm = \sin \theta^\pm \di \theta^\pm \di \phi^\pm$ are the solid angles of the final-state leptons in the respective
$W^{\pm}$ boson rest frames. The dependence on $m_{VV}$ and $\Theta$ is implicit. The matrix $\rho_{V_1V_2}$ describes the production of two spin-1 bosons, as in \eq{rho-qutrit}.

The polarization matrices $\Pi_\pm$ encode the chiral decay structure of the gauge bosons, with massless final leptons approximated as projectors. These are obtained by rotating the spin $\pm1$ states from the $z$-axis to arbitrary directions. In the Gell-Mann basis these are given by
\be
\Pi_\pm=\frac{1}{3}\, \mathbb{1}
+\frac{1}{2} \sum_{a=1}^8 \mathfrak{q}^a_\pm\, T^a \, , \label{pi_{f}}
\ee
where the Wigner functions $\mathfrak{q}^a_\pm$ are functions of the lepton spherical angles, their expressions can be found for instance in  \cite{Ashby-Pickering:2022umy,Barr:2024djo}.

The correlation coefficients $h_{ab}$ and polarization coefficients $f_a$, $g_a$ can be obtained from the single- and double-differential cross sections in \eq{x-sec-leptons} via the projections:
\bea
h_{ab} &=&  \frac{1}{4\, \sigma} \int \int \frac{\di \sigma}{\di\Omega^{+}\,\di\Omega^{-}} \, \mathfrak{p}_+^a \, \mathfrak{p}_-^b \,\di \Omega^+ \di \Omega^-\, .\label{hh}
\eea

\bea
f_{a} &=&  \frac{1}{2\, \sigma}  \int  \frac{\di \sigma}{\di\Omega^{+}} \, \mathfrak{p}_+^a \, \di \Omega^+ \, ,~~
g_{a}\,=\,  \frac{1}{2\, \sigma}  \int  \frac{\di \sigma}{\di\Omega^{-}} \, \mathfrak{p}_-^a \, \di \Omega^- \, . \label{ffgg}
\eea
Here, $\mathfrak{p}\pm^n$ are functions defined by
\be
\mathfrak{p}_\pm^n =   \sum_{m} (\mathfrak{m}^{-1}_{\pm})_{m}^{n} \,\mathfrak{q}_\pm^m \label{P_first}
\ee
and orthogonal to $\mathfrak{q}_{\pm}^m$ as
\be
\left(\dfrac{3}{4\, \pi} \right) \int \mathfrak{p}_\pm^n\, \mathfrak{q}_\pm^m \, \di \Omega^\pm = 2\,\delta^{nm} \, .
\ee
The matrix $\mathfrak{m}^{-1}_{\pm}$ is the inverse of the matrix $\mathfrak{m}_{\pm}$, whose matrix elements are defined as
\be
(\mathfrak{m}_{\pm})^{nm}= \left(\dfrac{3}{8\, \pi}  \right)\int \mathfrak{q}_\pm^n \, \mathfrak{q}_\pm^m \, \di \Omega^\pm \, .
\ee

For $Z$ boson decays, the polarization matrices $\Pi_\pm$ are not projectors due to their mixed chiral couplings in the neutral current Lagrangian
${\cal L}_Z=-i \frac{g}{\cos \theta_{W}} \Big[ g_{L} \gamma_{\mu}^L + g_{R} \gamma_{\mu}^R\Big] Z^{\mu}$, 
where (for charged leptons) $\gamma_{\mu}^{L/R}\equiv \gamma_{\mu}\frac{1}{2}(1\mp\gamma_5)$,   $g_{L} = -\tfrac{1}{2} + \sin^2 \theta_W$ and $g_{R} = \sin^2 \theta_W$. In this case, the functions $\mathfrak{q}_\pm^m$ are replaced by the weighted combination:
\be
\tilde{\mathfrak{q}}^{n} = \dfrac{1}{g_{R}^2 + g_{L}^{2}} \Big[ g_{R}^2 \, \mathfrak{q}^{n}_{+}+  g_{L}^2 \, \mathfrak{q}_{-}^{n}\Big]\, ,
\ee
with corresponding orthogonal functions
$\tilde{\mathfrak{p}}^{n} = \sum_{m}\mathfrak{a}^{n}_{m} \mathfrak{p}_+^m \, ,
\label{p_f}$
whose expression for $\mathfrak{a}^{n}_{m}$ can be found in \cite{Ashby-Pickering:2022umy,Barr:2024djo}. These functions are the same for both $\pm$ coordinate sets.
The projections in \eqs{hh}{ffgg} remain valid upon replacing $\mathfrak{p}_{\pm}^m$ with $\tilde{\mathfrak{p}}^n$. Furthermore, for identical bosons in $ZZ$ production, a symmetry factor of $1/2$ for $f_a$, $g_a$ and $1/4$ for $h_{ab}$ must be included~\cite{Ashby-Pickering:2022umy}.

\section{Entanglement and Bell inequality Violations at Colliders}
\label{sec:entanglement-Bell}
In this section, we present the main results concerning both the measurements and SM predictions of quantum entanglement and Bell inequality violations at colliders. We begin by reviewing the most relevant low-energy processes where quantum entanglement and Bell inequality violations have been already observed, focusing on neutral 
$B$-meson and charmonium decays into qubit and qutrit di-particle final states, as studied at the LHCb, Belle II, and BESIII experiments. We then discuss recent measurements of entanglement in top-quark pair production at the LHC, along with prospects for probing Bell inequality violations and quantum entanglement in diboson ($WW$, $ZZ$, and $WZ$) production, as well as via resonant Higgs-mediated channels ($W W^*$ and $Z Z^*$, with $*$ denoting an off-shell boson). Finally, we examine the potential for testing these quantum properties in $\tau$-lepton pair production at various collider facilities.

\subsection{$B^0$-meson decay in two Vector Mesons}
\label{Bdecays}

The first observation of Bell inequality violation at high energies--with a statistical significance well above the canonical $5\sigma$ threshold--was achieved in the context of neutral $B^0$ meson decays into two spin-1 vector mesons, $V_1$ and $V_2$~\cite{Fabbrichesi:2023idl,Gabrielli:2024kbz}. This result was obtained through the analysis of data from the LHCb collaboration~\cite{LHCb:2013vga}.

The theoretical framework employed to reconstruct the density matrix relies on the helicity amplitude spin formalism applied to final states involving two qutrits, as discussed in Section~\ref{sec:helicity-amplitude}.

In the specific case of a (pseudo)scalar meson decaying into two massive spin-1 particles, such as $B^0 \to V_1 V_2$, the expression for the density matrix in Eq.~\eqref{eq:rhoHA} simplifies significantly. Since the $B^0$ meson has spin zero, the density matrix becomes independent of the scattering angle $\theta$. In fact, the relevant Wigner $D$-matrix elements reduce to unity, and only three helicity amplitudes remain non-vanishing, as dictated by angular momentum conservation.

Due to helicity conservation, the (pure) quantum spin state of the two massive spin-1 particles can be expressed as\cite{Fabbrichesi:2023cev,Fabbrichesi:2023jep}
\bea
|\Psi \rangle &=& \frac{1}{\sqrt{|\mathcal{M}|^2}} \Big( w_{++}\, \ket{++} 
  +  w_{00} \, \ket{00} +  w_{--}\, \ket{--} \Big) \, ,
\label{pure}
\eea
where the normalization factor is given by
$|\mathcal{M}|^2 = |w_{++}|^2 + |w_{00}|^2 + |w_{--}|^2 $ .

The relative contributions of the transverse helicity states $\ket{++}$ and $\ket{--}$ compared to the longitudinal component $\ket{00}$ are dictated by angular momentum conservation. The corresponding helicity density-matrix $\rho = |\Psi\rangle \langle \Psi|$ is therefore entirely determined by the helicity amplitudes.
If one chooses the representation of the dimension-9 polarization basis as:
\begin{equation}
  \ket{+-} = 
  \begin{pmatrix}
   1 \\ 0 \\ 0\\ \vdots \\ 0 
  \end{pmatrix}\,,
  \quad
  \ket{+0} = 
  \begin{pmatrix}
   0 \\ 1 \\ 0\\ \vdots \\ 0 
  \end{pmatrix}\,,
  \quad
  \ket{++} = 
  \begin{pmatrix}
   0 \\ 0 \\ 1\\\vdots \\ 0 
  \end{pmatrix}\,,
  \quad \dots \quad,
  \ket{-+} = 
  \begin{pmatrix}
   0 \\ 0 \\ 0\\ \vdots \\ 1 
  \end{pmatrix}\, ,
\end{equation}
then the polarization density matrix takes the form:
\be
\small
\rho = \frac{1}{|\mathcal{M}|^2} \begin{pmatrix} 
  0 & 0 & 0 & 0 & 0 & 0 & 0 & 0 & 0  \\
  0 & 0 & 0 & 0 & 0 & 0 & 0 & 0 & 0  \\
  0 & 0 &  w_{++} w_{++}^* & 0 &  w_{++} w_{00}^* & 0 &  w_{++} w_{--}^* & 0 & 0  \\
  0 & 0 & 0 & 0 & 0 & 0 & 0 & 0 & 0  \\
  0 & 0 & w_{00} w_{++}^* & 0 & w_{00} w_{00}^* & 0 & w_{00} w_{--}^* & 0 & 0  \\
  0 & 0 & 0 & 0 & 0 & 0 & 0 & 0 & 0  \\
  0 & 0 & w_{--} w_{++}^* & 0 & w_{--} w_{00}^* & 0 & w_{--} w_{--}^* & 0 & 0  \\
  0 & 0 & 0 & 0 & 0 & 0 & 0 & 0 & 0  \\
  0 & 0 & 0 & 0 & 0 & 0 & 0 & 0 & 0  \\
\end{pmatrix} \, .
\label{rhoBVV}
\ee

The helicity amplitudes can be expressed in terms of the polarization amplitudes commonly adopted in experimental analyses~\cite{LHCb:2013vga,Belle:2005lvd,LHCb:2018hsm,ATLAS:2020lbz,LHCb:2023exl} through the following relations:
\be
\frac{w_{00}}{|\mathcal{M}|} = A_0 \,, \quad 
\frac{w_{++}}{|\mathcal{M}|} = \frac{A_{\parallel} + A_{\perp}}{\sqrt{2}} \,, \quad 
\frac{w_{--}}{|\mathcal{M}|} = \frac{A_{\parallel} - A_{\perp}}{\sqrt{2}} \, .
\ee
These relations allow us to rewrite the density matrix in Eq.~\eqref{rhoBVV} in terms of experimentally accessible polarization observables.

The decay channel with the most precisely measured polarization amplitudes is $B^0 \to J/\psi\, K^*$~\cite{LHCb:2013vga}. Assuming the density matrix takes the form given in Eq.~\eqref{rhoBVV}, it has been shown~\cite{Fabbrichesi:2023idl} that
\be\boxed{
  \mathscr{E} = 0.756 \pm 0.009 \,, \quad \text{and} \quad \mathcal{I}_3 = 2.548 \pm 0.015 } \,,
\ee
yielding a violation of the Bell inequality $\mathcal{I}_3 < 2$ with a statistical significance well beyond the $5\sigma$ threshold--numerically, at the $36\sigma$ level, as well as for the entropy of entanglement $\mathscr{E}>0$. 

To address the {\it locality loophole}--which concerns events not separated by a space-like interval (as in the case of $J/\psi\, K^*$ decays)--one must consider decay processes in which the two final-state particles are identical, such as $B_s \to \phi \phi$. In this case, since the two $\phi$ mesons have equal lifetimes, the decay exhibits an exponential time distribution with over 90\% of events occurring at space-like separations~\cite{Gabrielli:2024kbz}.

For the $B_s \to \phi \phi$ decay~\cite{LHCb:2023exl}, the corresponding analysis for the entropy of entanglement and Bell inequality violation ~\cite{Fabbrichesi:2023idl,Gabrielli:2024kbz} gives:
\be
\boxed{
\mathscr{E} = 0.734 \pm 0.037 \,, \quad \text{and} \quad \mathcal{I}_3 = 2.525 \pm 0.064 } \,,
\ee
which corresponds to a violation of the Bell inequality $\mathcal{I}_3 < 2$ at the $8.2\sigma$ level.

Quantum entanglement and Bell inequality violations have been also reported in other decay channels of neutral $B$ meson, but with minor statistical significance~\cite{Fabbrichesi:2023idl}.

\subsection{Charmonium Decays: Baryons and Vector Mesons}
Helicity amplitude measurements from charmonium decay experiments can also provide a means to quantify entanglement in the spin correlations of the final states and to probe potential violations of the Bell inequality.

Charmonium decays were initially identified in~\cite{Tornqvist:1980af,Tornqvist:1986pe} and \cite{Baranov:2008zzb,Baranov:2009zza}  as promising candidates for testing Bell inequality violations within particle physics. Specific decay channels such as  $\eta_c \to \Lambda \bar{\Lambda}$ and $J/\Psi\to \Lambda \bar{\Lambda}$ were first examined and later investigated in more detail in \cite{Chen:2013epa}.

Recently, Ref.~\cite{Fabbrichesi:2024rec} investigated quantum entanglement and violations of Bell inequalities in various bipartite final states resulting from the decays of spin-0, spin-1, and spin-2 charmonium systems. This analysis utilized helicity amplitude data provided by the BESIII Collaboration~\cite{BESIII:2018cnd,BESIII:2022qax,BESIII:2020fqg,BESIII:2024nif,BESIII:2021ypr,BESIII:2023drj,BESIII:2022lsz,BESIII:2023lkg,BESIII:2020lkm,BESIII:2023zcs,BESIII:2023euh}, as well as additional data on $J/\psi$ decays from the process $\Lambda_{b} \to J/\psi\, \Lambda$ as measured by the ATLAS~\cite{ATLAS:2014swk}, CMS~\cite{CMS:2018wjk}, and LHCb~\cite{LHCb:2013hzx} collaborations.
Below, we summarize the most significant findings from Ref.~\cite{Fabbrichesi:2024rec} concerning entanglement and Bell inequality violations in charmonium decays. Our focus here is on the most relevant results for spin-0 and spin-1 charmonium states. For a discussion of spin-2 states, which are of comparatively lower statistical significance, the reader is referred to Ref.~\cite{Fabbrichesi:2024rec}.

For a more comprehensive treatment of quantum correlations, decoherence effects, and the role of locality loopholes in charmonium decays, see Ref.~\cite{Fabbrichesi:2024rec}.

\subsubsection{Charmonium spin-0 states}
The scalar and pseudoscalar charmonium spin 0 states $\eta_c$ and  $\chi_c^0$ can decay into a pair of strange $\Lambda$ baryons and anti-baryons:
\be
\eta_c \to \Lambda + \bar \Lambda \quad \text{and} \quad \chi_c^0 \to \Lambda + \bar \Lambda ,
\ee
that have branching fractions of $(1.10\pm0.28)\times 10^{-3}$ and $(1.27\pm0.09)\times 10^{-4}$~\cite{ParticleDataGroup:2022pth}, respectively. These states are produced via the $e^+e^-$ production $e^+\, e^- \to J/\psi \to \gamma\, \eta_c$  and $ e^{+} \, e^{-} \to \psi (3686) \to \gamma \,\chi^{0} $.
The subsequent decays $\Lambda \to p\, \pi^-$ and $\bar \Lambda \to \bar p\, \pi^+$ enable to reconstruct the baryon polarizations through the angular distributions of the final particles states.

The associated quantum state is
$|\psi_0\rangle  \propto  w_{ \frac{1}{2}\,- \frac{1}{2}} \, |\tfrac{1}{2}, \tfrac{1}{2} \rangle \otimes |\tfrac{1}{2}, -\tfrac{1}{2} \rangle +
w_{-\frac{1}{2}\,\frac{1}{2} }\, |\tfrac{1}{2}, -\tfrac{1}{2} \rangle \otimes |\tfrac{1}{2}, \tfrac{1}{2} \rangle \label{state0}$
in which $w_{ij}$ are the helicity amplitudes and $|J,\, m\rangle$ the spin states, which gives the following density matrix
\be
\rho_{\Lambda\,\Lambda}  =  |\psi_0\rangle \langle \psi_0| = \frac{1}{2}\,  \begin{pmatrix} 
0&0&0&0\\
0& 1 & \pm1 &0\\
0&\pm 1&1 &0\\
0&0&0&0 
\end{pmatrix} \, .
\label{rho0}
\ee
The concurrence calculated from this density matrix is maximal, $\mathscr{C}[\rho] = 1$. Consequently, the Horodecki condition takes the value $\mathfrak{m}_{12} = 2$, indicating a maximal violation of the Bell inequality. While the evaluation does not require experimental input on the helicity amplitudes, uncertainties from the data analysis of these amplitudes are still essential to determine the statistical significance of the result. These uncertainties have not yet been provided by the BESIII experimental collaboration.

The next relevant case is the scalar state of the charmonium that can decay into a pair of spin-1 $\phi$ mesons
\be
\chi^{0}_c   \to \phi + \phi \,,
\ee
with branching fraction of $(8.48\pm0.26 \pm 0.27)\times 10^{-4}$~\cite{BESIII:2023zcs}. In total analogy to what is already discussed in the neutral $B$-meson decays into two spin-1 mesons  in Section~\ref{Bdecays}, the final state of the two $\phi$ mesons can  be written as 
$
|\Psi \rangle =w_{_{-1\, -1} }\, |-1,\, -1\rangle + w_{_{0\,0}}\, |0\, 0 \rangle+  w_{_{1\,1}}\, |1,\, 1\rangle  \, ,\label{pure}$
with
$|w_{_{-1\, -1}}|^2 + |w_{_{0\,0}}|^2 + |w_{_{1\,1}} |^2 =1$, and
$w_{_{1 \, 1}} = -w_{_{-1\, -1}}$ because of the conservation of parity. The corresponding polarization density matrix $\rho=|\Psi \rangle \langle \Psi|$ is analogous to the one already provided in Eq.~\eqref{rhoBVV}.

Based on the analysis of helicity amplitude data presented in~\cite{BESIII:2023zcs}, the entanglement entropy is found to be~\cite{Fabbrichesi:2024rec}
\be
\boxed{
\mathscr{E}[\rho] = 0.531 \pm 0.040}\, ,
\ee
which deviates from the null hypothesis with a significance of $13.3\sigma$.
After optimization, the expectation value of the Bell operator is~\cite{Fabbrichesi:2024rec}
\be
\boxed{
\mathcal{I}_3= 2.296 \pm 0.034}\, ,
\ee
providing a clear test of Bell inequality violation in the two qutrit systems in the charmonium decay, with a significance of $8.8\sigma $.

\subsubsection{Charmonium spin-1 states}
The decay of a spin-1 particle exhibits a dependence on the scattering angle $\theta$ through the structure of its polarization density matrix, as shown by the decomposition in terms of helicity amplitudes in \eq{eq:rhoHA}.
Further details on the corresponding polarization density matrices for various final states in spin-1 charmonium decays can be found in\cite{Fabbrichesi:2024rec}.
As a result, the quantification of entanglement and the violation of Bell inequalities acquire an angular dependence that can be exploited to identify regions where these effects are enhanced.

The following decays of spin-1 charmonium states have been analyzed in~\cite{Fabbrichesi:2024rec}, using data from various BESIII Collaboration analyses. In several of these channels, the violation of the Bell inequality is observed with a significance exceeding $5\sigma$
\bea
J/\psi  &\to&  \Lambda +\bar \Lambda ~\text{\cite{BESIII:2022qax}}\, ,
\nonumber \\
J/\psi & \to & \Xi^- +\bar \Xi^+~\text{\cite{BESIII:2021ypr}}\, , \, \Xi^0 +\bar \Xi^0 ~\text{\cite{BESIII:2023drj}}\, ,
\nonumber \\
J/\psi & \to & \Sigma^{-} +\bar \Sigma^{+}~\text{\cite{BESIII:2020fqg}}\, , \,  \Sigma^{0} +\bar \Sigma^{0}~\text{\cite{BESIII:2024nif}}, \nonumber \\
\psi(3686)& \to &   \Xi^- +\bar \Xi^+~\text{\cite{BESIII:2022lsz}}\, , \,  \Sigma^- +\bar \Sigma^+\text{\cite{BESIII:2020fqg}}\, , \Sigma^{0} +\bar \Sigma^{0}~\text{\cite{BESIII:2024nif}}, \nonumber \\
 \chi_{c}^{1} & \to & \phi+ \phi~\text{\cite{BESIII:2023zcs}}\, .\nonumber  
 \eea
 This includes the decays $\psi(3686) \to \Lambda + \bar{\Lambda}$\cite{BESIII:2023euh} and $\psi(3686) \to \Xi^0 + \bar{\Xi}^0$\cite{BESIII:2023lkg}, where the violation is present but with lower statistical significance. In contrast, the decay $\psi(3686) \to \Omega^- + \Omega^+$ ~\cite{BESIII:2020lkm} shows evidence of quantum entanglement, though without a significant violation of the Bell inequality.

 In Table~\ref{charmonium-spin1}, we report a summary of the corresponding results for the quantum entanglement and violation of the Bell inequality--quantified for qubits by $\mathscr{C}[\rho]$ and $\mathfrak{m}_{12}$ observables respectively--in the most sensitive charmonium decay channels into spin-$\tfrac{1}{2}$ baryons~\cite{Fabbrichesi:2024rec}. The results are evaluated at $\theta = \pi/2$, where the violation reaches its maximum and the degree of entanglement is also maximal.

\begin{table}[h!]
\tablestyle[sansboldbw]
\begin{tabular}{*{4}{p{0.2\textwidth}}}
\theadstart
    \thead  decay &\thead $\mathscr{C}[\rho]$  &\thead $\mathfrak{m}_{12}$ &
    \thead significance \\
\tbody
 $J/\psi \to \Lambda \bar \Lambda$ & 0.475 $\pm$ 0.004 & 1.225 $\pm$ 0.004 & \hskip1cm 56.3 \\
 $\psi(3686) \to \Lambda \bar \Lambda$ & 0.69 $\pm$ 0.07    &1.476 $\pm$ 0.10   & \hskip1cm 4.76 \\
 $J/\psi \to \Xi^- \bar \Xi^+$     &  0.586 $\pm$ 0.016  &1.343 $\pm$ 0.018  &  \hskip1cm 19.1 \\
$J/\psi \to \Xi^0 \bar \Xi^0$  &   0.514 $\pm$ 0.016  &1.264  $\pm$ 0.017  &  \hskip1cm 15.6 \\
$\psi(3686) \to\Xi^- \bar \Xi^+$ & 0.693 $\pm$ 0.068 &1.480 $\pm$ 0.095  & \hskip1cm 5.05\\
$\psi(3686) \to\Xi^0 \bar \Xi^0$  & 0.665 $\pm$ 0.119  &1.442 $\pm$ 0.161 &  \hskip1cm 2.75 \\
$J/\psi \to \Sigma^- \bar \Sigma^+$ &  0.508 $\pm$ 0.007  &1.258 $\pm$ 0.007 & \hskip1cm 36.9 \\
$\psi(3686) \to\Sigma^- \bar \Sigma^+$ &0.682 $\pm$ 0.032 &1.465 $\pm$ 0.043  &  \hskip1cm 10.8 \\
$J/\psi \to \Sigma^0 \bar \Sigma^0$ &   0.413 $\pm$ 0.009  &1.171 $\pm$ 0.007   & \hskip1cm 24.4\\
$\psi(3686) \to\Sigma^0 \bar \Sigma^0$ &  0.814 $\pm$ 0.040  &1.663 $\pm$ 0.065   & \hskip1cm 10.2 \\
   \hline%
  \tend
\end{tabular}
\caption{\footnotesize \textrm{Summary of quantum entanglement ($\mathscr{C}[\rho]$) and Bell inequality violation ($\mathfrak{m}_{12}$) in spin 1 charmonium states decaying into a pair of baryons (two qubits), the significance refers only to the Bell inequality violation~\cite{Fabbrichesi:2024rec}.}}
 \label{charmonium-spin1}
\end{table}

Finally, regarding the violation of the Bell inequality in the charmonium decay into two spin-1 mesons (two qutrits), $\chi^1_{c} \to \phi + \phi$, using the helicity amplitude data reported in~\cite{BESIII:2023zcs}, the following result is obtained at $\theta = \pi/2$~\cite{Fabbrichesi:2024rec}:
\be
\boxed{
 \mathcal{I}_3 = 2.296 \pm 0.003}\, .
 \ee
The significance of this Bell inequality violation is found to be $98.7\sigma$, although it should be noted that this estimate accounts only for statistical uncertainties.
Concerning entanglement, being a bipartite system of two qutrits produced in a mixed state, only the quantity $\mathscr{C}_2$, defined in \eq{C_2}, can be used. This provides a lower bound on entanglement, giving at $\theta = \pi/2$ its maximum value $\mathscr{C}_2 \simeq 1$ as reported in ~\cite{Fabbrichesi:2024rec}.

\subsection{Top-quark pair production}
\label{sec:top}
At the parton level, top-quark pair production ($t\bar{t}$) at the LHC proceeds through two main processes t the tree-level: quark–antiquark annihilation ($q\bar{q} \to t\bar{t}$) and gluon–gluon fusion ($gg \to t\bar{t}$). The corresponding SM Feynman diagrams are shown in Fig.~\ref{fig:ttbar}. Therefore, top-quark pairs are produced in a mixed state, arising from the different initial states.

 \begin{figure}[h!]
\begin{center}
\includegraphics[width=5.5in]{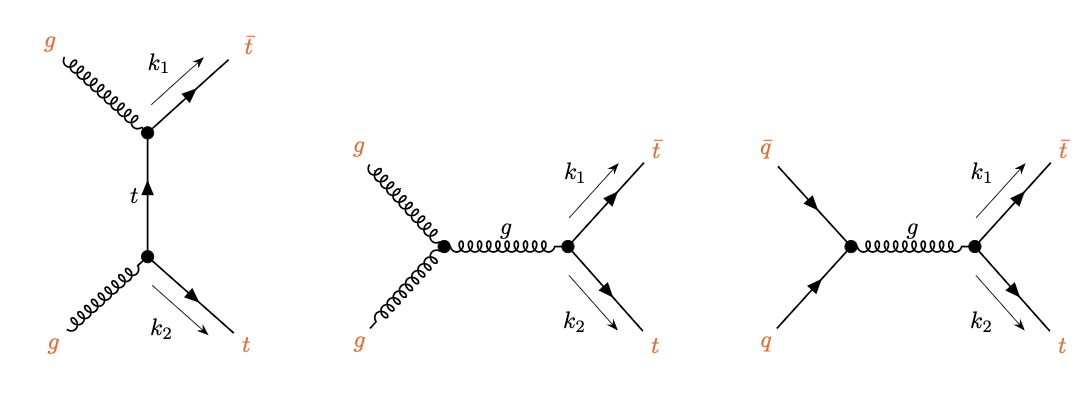}
\caption{\small \label{fig:ttbar}  Feynman diagrams (at partonic tree-level)  for top-antitop ($t\bar t$) production, for  gluon ($gg$) and quark-pair ($q\bar{q}$) initial states. (Figure from \cite{Barr:2024djo} under the \href{https://creativecommons.org/licenses/by/4.0/}{CC BY 4.0} licence).
}
\end{center}
\end{figure}

The unpolarized differential cross section for the process
\be
p + p \to t + \bar{t}
\label{eq:pptt}
\ee
can be expressed as~\cite{Bernreuther:2001rq,Afik:2020onf,Fabbrichesi:2021npl}:
\be
\frac{\di \sigma}{\di \Omega\, \di m_{t\bar t}} = \frac{\alpha_s^2 \beta_t}{64 \pi^2  m_{t\bar t}^2} \left\{ L^{gg} (\tau) \, \tilde A^{gg}[m_{t\bar t},\, \Theta]+L^{qq} (\tau)\, \tilde A^{qq}[m_{t\bar t},\, \Theta]  \right\},
\label{eq:x-sec-tt}
\ee
where the contributions from the partonic processes $gg \to t\bar{t}$ and $q\bar{q} \to t\bar{t}$ (see Fig.~\ref{fig:ttbar}) are weighted by the respective parton luminosities $L^{gg}(\tau)$ and $L^{qq}(\tau)$:
\be
L^{gg} (\tau)= \frac{2 \tau}{\sqrt{s}} \int_\tau^{1/\tau} \frac{\di z}{z} q_{g} (\tau z) q_{g} \left( \frac{\tau}{z}\right),\quad
L^{qq} (\tau)= \sum_{q=u,d,s}\frac{4 \tau}{\sqrt{s}} \int_\tau^{1/\tau} \frac{\di z}{z} q_{q} (\tau z) q_{\bar q} \left( \frac{\tau}{z}\right),
\label{eq:lumPDF}
\ee
where $q_j(x)$ are the parton distribution functions (PDFs), $\alpha_s = g^2 / (4\pi)$ the strong fine-structure constant, and $\tau = m_{t\bar{t}} / \sqrt{s}$ with $m_{t\bar{t}}$ the invariant mass of the top-quark pair. Explicit expressions for the quantities $\tilde A^{gg}$ and $\tilde A^{qq}$ can be found, for instance, in \cite{Barr:2024djo}. Their numerical evaluation can be based on recent PDF sets, such as {\sc PDF4LHC21}~\cite{PDF4LHCWorkingGroup:2022cjn}, for $\sqrt{s} = 13$~TeV and factorization scale of the order of $m_{t\bar{t}}$.

The correlation coefficients $C_{ij}$ in the polarization density matrix for $t\bar{t}$ production, in the basis $\{{\bf {\hat{r}, \hat{n}, \hat{k}}}\}$ defined in \eq{basis-rnk} in section \ref{sec:simulation-qubits}, have been computed in~\cite{Bernreuther:2001rq,Afik:2020onf,Fabbrichesi:2021npl} and are given by
\be
C_{ij} [m_{t\bar t},\, \Theta]= \frac{L^{gg} (\tau)\, \tilde C_{ij}^{gg}[m_{t\bar t},\, \Theta]+L^{qq} (\tau)\, \tilde C_{ij}^{qq}[m_{t\bar t},\, \Theta]+\Big\{\Theta\to \Theta+\pi\Big\}} {L^{gg}(\tau) \, \tilde A^{gg}[m_{t\bar t},\, \Theta]+L^{qq} (\tau)\, \tilde A^{qq}[m_{t\bar t},\, \Theta]+\Big\{\Theta\to \Theta+\pi\Big\}}\, .
\label{eq:Cij-top}
\ee
The last terms involving the exchange $(\Theta\to \Theta + \pi)$ in \eq{eq:Cij-top} account for the symmetry under the exchange of quarks and antiquarks originating from either proton beam at the LHC, as both configurations share the same parton luminosity functions.
In the SM, the single-spin polarization coefficients for quark-initiated processes vanish identically at leading order, i.e., $B_i^{q\bar{q}} = 0$, except for the small contributions induced by the higher-order electroweak corrections or due to the $b\bar{b}\to t\bar{t}$ contribution--basically the interference of the $W$-mediated ($t$-channel) diagram in $b\bar{b}\to t\bar{t}$ with the corresponding gluon-mediated ($s$-channel) one.

The explicit expressions for $\tilde C_{ij}^{gg}$ and $\tilde C_{ij}^{qq}$ can be found in \cite{Barr:2024djo,Bernreuther:2001rq,Afik:2020onf,Fabbrichesi:2021npl}. These are related to the partonic correlation coefficients $C_{ij}^{gg}$ and $C_{ij}^{qq}$ through the relations:
\be
\tilde{C}_{ij}^{gg}= C_{ij}^{gg} \, \tilde{A}^{gg}, \qquad \tilde{C}_{ij}^{qq}= C_{ij}^{qq} \, \tilde{A}^{qq}.
\ee

\subsubsection{Entanglement}
Top-quark pair production has been the first process analyzed in the current series of studies on quantum entanglement at high energy. In~\cite{Afik:2020onf}, the expected entries of the polarization density matrix were evaluated in the reference frame introduced in~\cite{Bernreuther:2010ny} and the concurrence  $\mathscr{C}[\rho]$ computed.

The dependence of the entries $C_{ij}$ of the polarization density matrix in \eq{eq:Cij-top} on the scattering angle $\Theta$ and velocity factor $\beta_t = \sqrt{1 - 4m_t^2 / m_{t\bar{t}}^2}$ in the $t\bar{t}$ center of mass, significantly simplify in the central region at $\Theta = \pi/2$, where the top-quark pair is produced transversely and the entanglement reaches its maximum \cite{Barr:2024djo,Fabbrichesi:2021npl}. There are two regions where entanglement is particularly significant: a narrow region near threshold, and the boosted regime for scattering angles close to $\pi/2$. In the latter case, the violation of the Bell inequality (quantified by $\mathfrak{m}_{12}$) also reaches its maximum. See the contour plots of the observables $\mathscr{C}[\rho]$ and $\mathfrak{m}_{12}$ as functions of $m_{t\bar{t}}$ and $\cos{\Theta}$ in Fig.~\ref{fig:topm1m2}.

In \cite{Afik:2020onf} it was shown that  entanglement is very sensitive to the kinematic variables. This can be understood by noticing that the quantity
\be
D= \frac{1}{3} \, \Tr C_{ij}
\ee
close to $t\bar{t}$ threshold is expected to be smaller than $-1/3$. This is a sufficient condition for  entanglement, since the factor $D$ is directly connected to the concurrence by the relation $\mathscr{C}[\rho]=\max [-1-3D,\,0]/2$~\cite{Afik:2020onf}.

The coefficient $D$ can be extracted from data by analyzing the differential cross section for the process $p+p \to t+\bar t \to \ell^+ \ell^- + \text{jets}+ E^{\rm miss}_{\rm T}$ (with
 $E^{\rm miss}_{\rm T}$ the transverse missing energy from neutrinos) 
\be
\frac{1}{\sigma} \frac{\di \sigma}{\di \cos \phi} = \frac{1}{2} \Big( 1 - D \cos \phi \Big)\, ,
\label{dsigmaD}
\ee
where $\phi$ is the angle between the respective leptons as computed in the rest frame of the decaying top and anti-top~\cite{Afik:2020onf}.

The ATLAS~\cite{ATLAS:2023fsd} and CMS~\cite{CMS:2024pts,CMS:2024zkc} Collaborations have  recently analyzed the $13\, {\rm TeV}$ $pp$ data and extracted the value of $D$ from the differential cross section in \eq{dsigmaD}.
The analysis focused on fully leptonic top quark pair events, identified by tagging two oppositely charged leptons with high transverse momentum. The observable $D$ was measured at the particle level in the near-threshold region--specifically, within the ranges $340\, {\rm GeV} < m_{t\bar{t}}< 380\, {\rm GeV}$ for ATLAS and $345\, {\rm GeV} < m_{t\bar{t}} < 400\, {\rm GeV}$ for CMS. The measurements yielded:
\be
D=\left\{
\begin{array}{l}
-0.537 \pm 0.002\; [{\rm stat.}] \pm 0.019\; [{\rm syst.}]~~(\text{ATLAS}) \text{\cite{ATLAS:2023fsd}}\\\\
  -0.480  ~^{+0.026}_{-0.029}~~~~(\text{CMS}) \text{\cite{CMS:2024pts,CMS:2024zkc}}
\end{array}
\right.
\ee 
The measured $D$ value is in both experiments smaller than $-1/3$ with an observed significance of more than $5\sigma$ for ATLAS and CMS collaborations, thus providing the first experimental observation of the presence of entanglement between the spins of the top quarks.\footnote{In \cite{Fabbrichesi:2025psr} it has been recognized that the observed entanglement is {\it local} in the energy region near the production threshold--with no violation of Bell inequality expected--while, in contrast, {\it non-local} entanglement would be observed in the central boosted regions with $m_{t\bar{t}} > 800$.}

The observed entanglement exceeds that predicted by simulations, suggesting that the latter may require more accurate modeling of near-threshold effects in $t\bar{t}$ production.
In the CMS analysis, the Monte Carlo simulations included the colour-singlet contribution from toponium bound states—a pseudoscalar $t\bar{t}$ bound state—as calculated in~\cite{Fuks:2021xje}. The inclusion of this contribution tends to enhance the predicted entanglement and improves agreement between simulation and data.

It is worth noting that more recently in~\cite{Nason:2025hix}, resummation of all corrections scaling as powers of $\alpha_S/v$--where $v$ is the velocity of the top quark in the $t\bar{t}$ rest frame--has been incorporated into observables sensitive to spin correlations in $t\bar{t}$ production. These corrections, dominated by values of $v$ of order $\alpha_S$, are analogous to the Sommerfeld-Fermi corrections of QED~\cite{Sommerfeld:1931qaf,Fermi:1934hr}, accounting for the non-perturbative QCD-like Coulombic effects. In~\cite{Nason:2025hix} it is shown that their inclusion significantly reduces, or even eliminates, the tension between theoretical predictions and experimental data.

 \begin{figure}[h!]
\begin{center} 
 \includegraphics[width=3.2in]{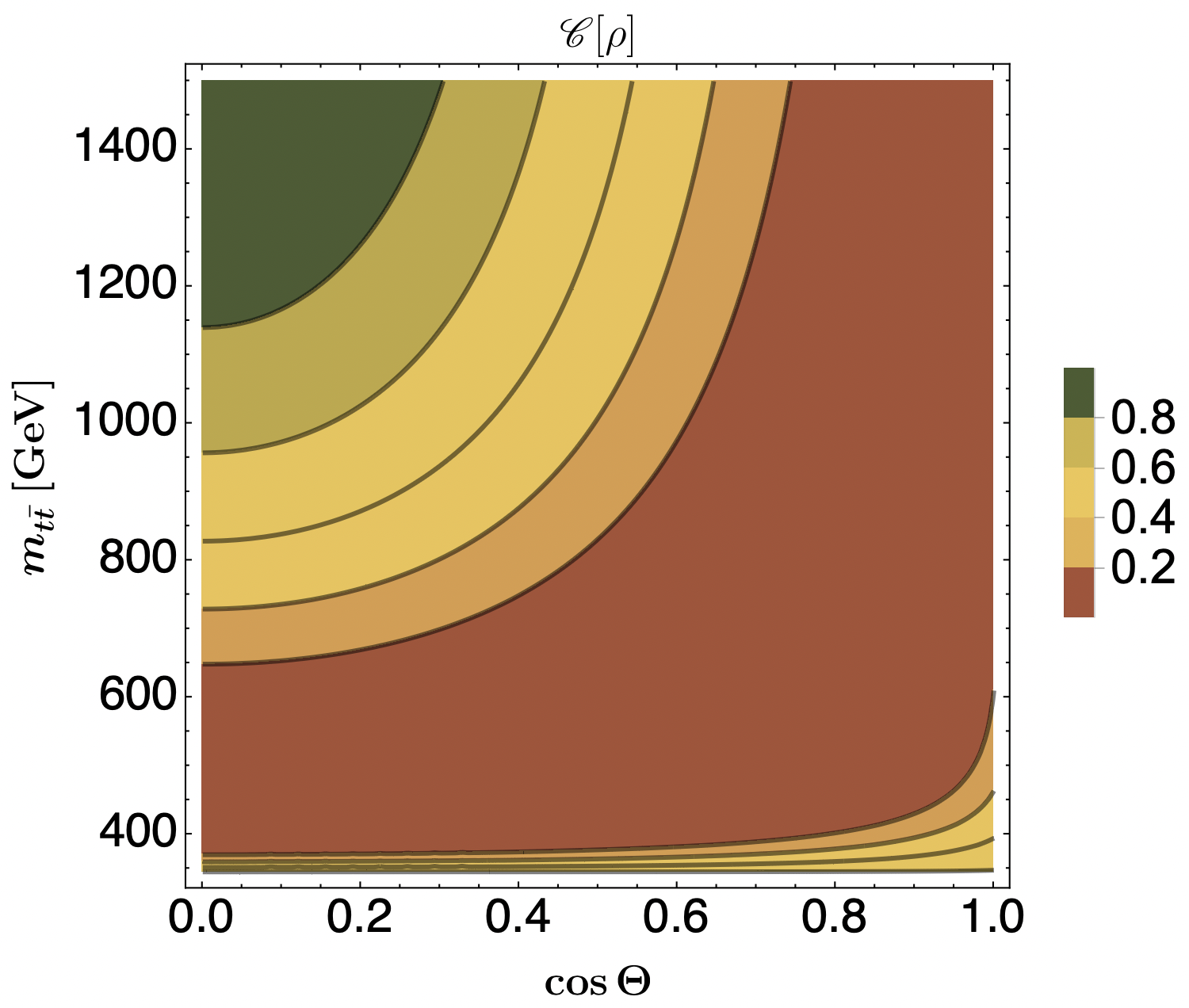}
  \includegraphics[width=3.25in]{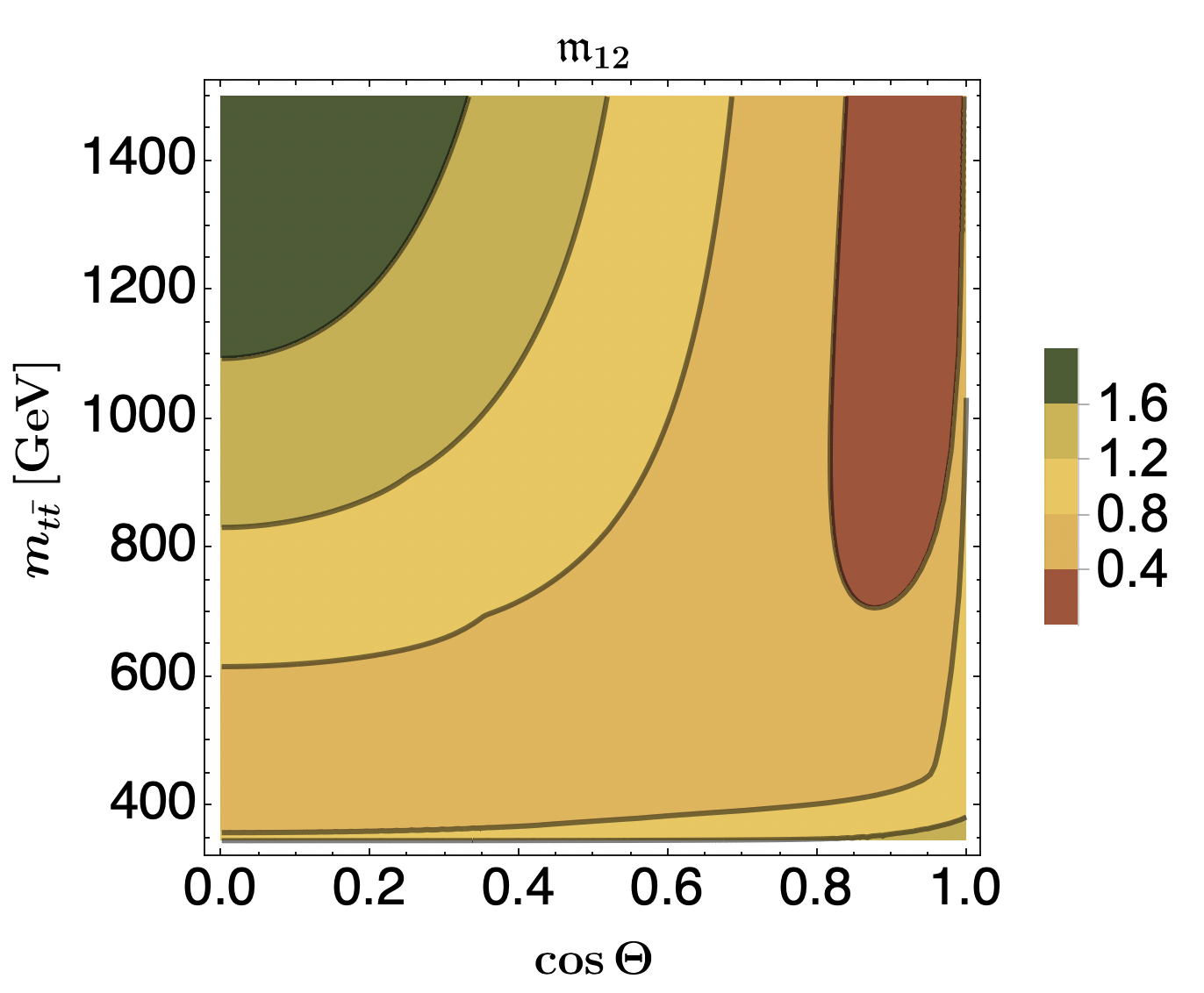}
  \caption{\small  The observables $\mathscr{C}[\rho]$ (contour plot on the left) and $\mathfrak{m}_{12}$ (contour plot on the right) for $t\bar t$ production as  functions of the kinematic variables $\Theta$ and $m_{t \bar t}$ across the entire available space (they  are symmetric for $\cos \Theta<0$). (Figures from \cite{Barr:2024djo} and revised from \cite{Fabbrichesi:2022ovb} under the
    \href{https://creativecommons.org/licenses/by/4.0/}{CC BY 4.0} licence). 
  \label{fig:topm1m2} 
}
\end{center}
\end{figure}

\subsubsection{Bell inequality violation}
\label{sec:top-bell}

The entanglement between the top–antitop quark pair leads to a measurable violation of the Bell inequality, as quantified via the Horodecki criterion~(\ref{eigenvalue-inequality})~\cite{Fabbrichesi:2021npl}. Specifically, this violation is captured by the observable
\be
\mathfrak{m}_{12} \equiv m_1 + m_2 > 1,
\label{eq:m12}
\ee
as introduced in Section~\ref{sec:qubits}, where $m_{1,2}$ are the two largest eigenvalues of the $CC^T$ matrix in \eq{eigenvalue-inequality}.
The distribution of $\mathfrak{m}_{12}$ for the top-pair production at the LHC throughout the accessible kinematic phase space is illustrated on the right panel of Fig.~\ref{fig:topm1m2}.

As shown in Fig.~\ref{fig:topm1m2}, both quantum entanglement and the violation of the Bell inequality--encoded in the value of $\mathfrak{m}_{12}[C]$--are enhanced at larger scattering angles and higher invariant masses $m_{t\bar{t}}$. In line with the qualitative expectations outlined in the previous section, the region where $\mathfrak{m}_{12}$ is most prominent corresponds to $m_{t\bar{t}} > 900~\text{GeV}$ and $\cos\Theta/\pi < 0.2$. Within this kinematic bin, the mean value of $\mathfrak{m}_{12}$ reaches 1.44~\cite{Fabbrichesi:2022ovb}.

Monte Carlo simulations of quantum observables in top-quark pair production--considering both fully and semi-leptonic decay channels--have been performed in \cite{Fabbrichesi:2021npl,Severi:2021cnj,Aguilar-Saavedra:2022uye,Dong:2023xiw,Han:2023fci}. These studies successfully reproduce the SM analytic predictions and also provide estimates of the associated uncertainties for entanglement and Bell inequality violation.

In \cite{Fabbrichesi:2021npl}, the dileptonic process
$p+p \to t+\bar t \to \ell^\pm \ell^\mp + \text{jets}+ E^{\rm miss}_{\rm T}$
is simulated at leading order using \textsc{MadGraph5\_aMC@NLO}~\cite{Alwall:2014hca}, followed by parton showering and hadronization~\cite{Sjostrand:2014zea}, and detector-level reconstruction with the ATLAS detector card~\cite{deFavereau:2013fsa}.

From the simulated events, as explained in section \ref{sec:simulation-qubits}, angular correlations between the charged leptons--specifically, the product $\cos \theta^i_+ \cos \theta^j_-$--are used to extract the matrix $C_{ij}$, which encodes entanglement and Bell inequality observables. This requires reconstructing the top rest frame, including the neutrino momenta. In \cite{Fabbrichesi:2021npl}, the value of  $\mathfrak{m}_{12}$ has been estimated after correcting for the bias. The predicted significance of Bell inequality violation reaches $3\sigma$ for the combined Run~1 and Run~2 datasets at the LHC (with 300~fb$^{-1}$ of integrated luminosity), while this increases to $4\sigma$ at the high-luminosity LHC (Hi-Lumi LHC) with 3~ab$^{-1}$~\cite{Fabbrichesi:2021npl}. In contrast, a lower significance is reported in \cite{Severi:2021cnj} for the same kinematic region: below $1\sigma$ for Run~1 plus Run~2, and only 1.8$\sigma$ at the Hi-Lumi LHC.
An increase of a factor 1.6 in significance is expected between the fully leptonic and the semi-leptonic channels, as shown in the analysis of \cite{Dong:2023xiw,Han:2023fci}.

In addition to the Bell non-locality, there are also other studies that have analyzed quantum discord~\cite{Han:2024ugl,Afik:2022dgh}, steerability~\cite{Afik:2022dgh}, and non-negative conditional entropy~\cite{Han:2024ugl} in the $t\bar{t}$ final state system. These investigations go beyond the traditional focus on Bell inequalities and provide complementary insights into the structure of quantum correlations in $t\bar{t}$.
 Taken together, these works illustrate that the $t\bar{t}$ 
  system constitutes a rich laboratory for testing different aspects of quantum information theory, broadening our understanding of how quantum correlations manifest in particle physics experiments.

\subsubsection{Probing new physics with entanglement}
\label{sec:top-new-physics}

Entanglement can also be used to probe sensitivity to new physics contributions beyond the SM. In this case the advantage with respect to usual observables as unpolarized as cross sections rely on the fact that entanglement is sensitive to all polarization amplitudes which are encapsulated into the polarization density matrix. On the other hand, the uncertainties on the measurements of the polarizations can reduce this sensitivity.

Modifications of the entanglement for the spins of the top-quark pairs in this framework has been studied in~\cite{Aoude:2022imd}. In this approach the effective Lagrangian is given by the SM effective theory (SMEFT) which reads \cite{Grzadkowski:2010es}
\be
{\cal L}{\text{\tiny SMEFT}} = {\cal L}{\text{\tiny SM}} + \frac{1}{\Lambda^2} \sum_i c_i {\cal O}_{i},, \label{L_smeft}
\ee
where $\Lambda$ denotes the scale of new physics and $c_i$ the Wilson coefficients. At leading order in QCD, all $CP$-even dimension-six operators are included in the analysis of top-spin entanglement: one with zero fermions, two with two fermions, and fourteen with four fermions~\cite{Aoude:2022imd}. The contribution of higher-order terms in the SMFET expansion have been analyzed in \cite{Severi:2022qjy}.

The concurrence is affected by interference terms $\propto c_i/\Lambda^2$, linear in the cross section, and by quadratic terms $\propto (c_i/\Lambda^2)^2$. Near threshold, interference effects are small while quadratic terms reduce the concurrence. At high energies, both contributions suppress it significantly.

The use of entanglement to constrain the new physics contribution to the gluon magnetic-like dipole operator has been analyzed in~\cite{Fabbrichesi:2022ovb}. At this aim the effective Lagrangian is adopted
\be
{\cal L}_{\text{\tiny  dipole}}=\frac{c_{\,t\G}}{\Lambda^2} \big( {\cal O}_{t\G} +{\cal O}_{t\G}^\dag \big) \quad \text{with}  \quad
 {\cal O}_{t\G} =g_s \left(\bar Q_L \, \sigma^{\mu \nu} \, T^a\, t_R \right) \tilde{H}  G_{\mu\nu}^a \, . \label{L_dip}\, ,
 \ee
 where $Q_L$ and $t_R$ stands for the $SU(2)_L$ left-handed doublet of top-bottom quarks and right-handed top quark fields respectively, while $\tilde{H}$ is as usual the dual of the $SU(2)_L$  doublet Higgs field, with $v$ the SM Higgs vacuum expectation value.  The chromo-magnetic-like dipole moment is then given by  $ \mu = - \frac{\sqrt{2} m_t v}{\Lambda^2} c_{\,tG}\ $.
A simple Monte Carlo study in~\cite{Fabbrichesi:2022ovb} shows that, in the kinematic region $m_{t\bar t} > 900$ GeV and $2\Theta/\pi > 0.85$ (where SM and new-physics predictions differ by about 3\%), a $2.3\sigma$ separation can be achieved for $\mu=0.003$ using Run2 LHC data. This agrees with the analysis of\cite{Aoude:2022imd} (for $c_{t\G}=-0.1$ at $\Lambda=1$ TeV) and improves upon current bounds of $\mu \simeq 0.02$~\cite{CMS:2019kzp}.

\subsection{Diboson production at LHC and linear colliders}
The potential for measuring the quantum observables in the qutrits system characterized by the gauge diboson production at the LHC has been explored in \cite{Ashby-Pickering:2022umy,Fabbrichesi:2023cev,Aoude:2023hxv} by looking at the processes
\be
pp \to W^+W^-,~ ZZ,~  WZ\, .
\ee
These final states can be generated through electroweak processes across a continuous spectrum of diboson invariant masses. The relevant Feynman diagrams at the tree-level parton-level are illustrated in  illustrated in Fig.~\ref{fig:DYVV}.

In \cite{Fabbrichesi:2023cev}, the polarization density matrix of the diboson system has been derived analytically within the SM framework, following the method explained in section \ref{sec:rho-spin1}.
In contrast, \cite{Ashby-Pickering:2022umy} reconstructs the same density matrix through event simulations of diboson decays, employing the method described in Section \ref{sec:simulation-qutrits}.

The polarization density matrix of the diboson system can be constructed using the decomposition in the Gell-Mann basis, as outlined in \eq{rho-qutrit}.
In particular, following the analogy with \eq{eq:Cij-top} for the qubit system in top-pair production, we obtain the correlation coefficients $h_{ab}$ as
\be
h_{ab}[\mVV, \Theta] = \frac{\sum_{q=u,d,s} L^{q\bar q}(\tau)
\left( \tilde{h}^{q \bar q}_{ab}[\mVV, \Theta] +\Big\{\Theta\to \Theta+\pi\Big\}
\right)}
{\sum_{q=u,d,s} L^{q\bar q}(\tau)\left( A^{q \bar q}[\mVV, \Theta] +\Big\{\Theta\to \Theta+\pi\Big\} \right)}\, ,
\label{eq:habDY}
\ee
where $\{a,b\}\in\{1, \dots, 8\}$, $\mVV$ denotes the invariant mass of the diboson final state and $\Theta$ is the scattering angle in their center-of-mass frame. The shorthand $A^{q \bar q}$ represents the spin-summed squared amplitude of the process, while $\tilde{h}_{ab} = A^{q \bar q} h_{ab}$. The functions $L^{q\bar q}(\tau)$ are the quark parton luminosities defined in \eq{eq:lumPDF}. Analogous expressions holds for the polarization coefficients $f_a$ and $g_a$. 
The functions $h_{ab},f_a,g_a$ have been computed in \cite{Fabbrichesi:2023cev} and their  analytical expressions can be found in~\cite{Fabbrichesi:2023cev,Barr:2024djo}.
\begin{figure}[h!]
  \begin{center}
  \includegraphics[width=5in]{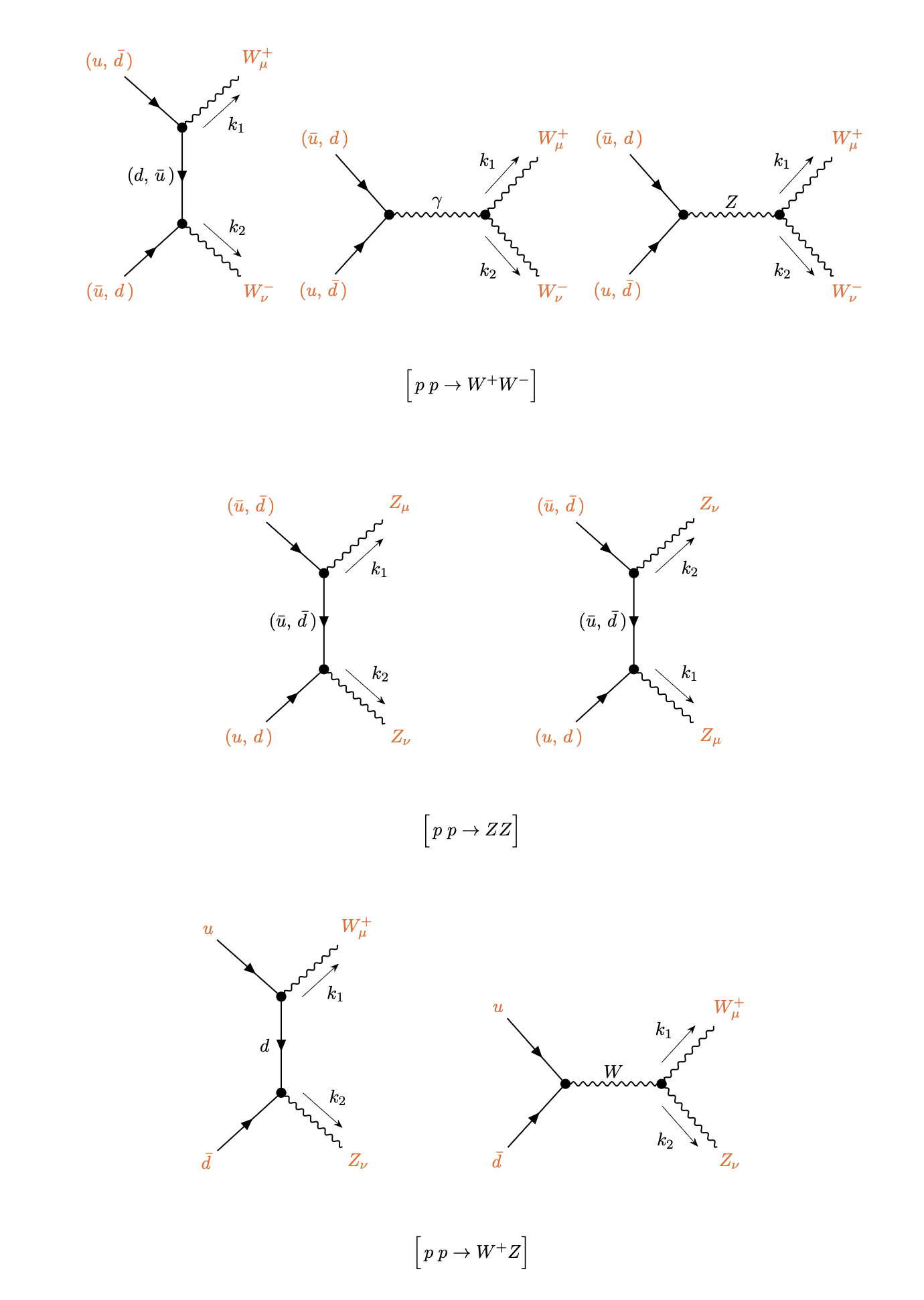}
  \caption{\small Feynman diagrams for the parton-level processes $pp \to W^+W^-$ (top), $pp \to ZZ$ (middle), and $pp \to W^+Z$ (bottom) involving first-generation quarks. Higgs-mediated diagrams are omitted in the massless quark limit. Arrows on fermion lines indicate momentum flow.
(Figure from \cite{Barr:2024djo} under the \href{https://creativecommons.org/licenses/by/4.0/}{CC BY 4.0} licence).
  \label{fig:DYVV} 
  }
  \end{center}
  \end{figure}

Quantum entanglement and the violation of Bell inequalities for qutrits are characterized by the observables $\mathscr{C}_2$ and ${\cal I}_3$, as defined in \eq{C2} and \eq{I3}, respectively. In the case of ${\cal I}_3$ a maximization procedure is applied \cite{Fabbrichesi:2023cev}, as discussed in section \ref{sec:qutrits}.

In Figs. \ref{fig:WW} and \ref{fig:ZZ}, we present the contour plots of the observables $\mathscr{C}_2$ and ${\cal I}_3$ in the $(\mVV, \cos{\Theta})$ plane for $WW$ and $ZZ$ production, respectively, as obtained in \cite{Fabbrichesi:2023cev}. Results for $WZ$ production are not shown, since the analysis in \cite{Fabbrichesi:2023cev} found no significant deviation from the null hypothesis for Bell inequality violation across the relevant kinematic region--although the presence of entanglement remains a possibility to explore \cite{Fabbrichesi:2023cev,Ashby-Pickering:2022umy}.

Monte Carlo simulations of diboson production at the LHC were performed in \cite{Ashby-Pickering:2022umy, Bernal:2023ruk} using \textsc{MadGraph5\_aMC@NLO}~\cite{Alwall:2014hca}, incorporating spin correlations, relativistic effects, and Breit–Wigner propagators. Events were generated at leading order for a 13 TeV center-of-mass energy, focusing on 4-lepton final states. In line with analytic predictions, entanglement is expected at large scattering angles and invariant masses above 400 GeV in both $WW$ and $ZZ$ channels. However, no significant Bell inequality violation is expected--even at high luminosity (3 ab$^{-1}$)--once statistical uncertainties are taken into account.
\begin{figure}[h!]
  \begin{center}
  \includegraphics[width=3.2in]{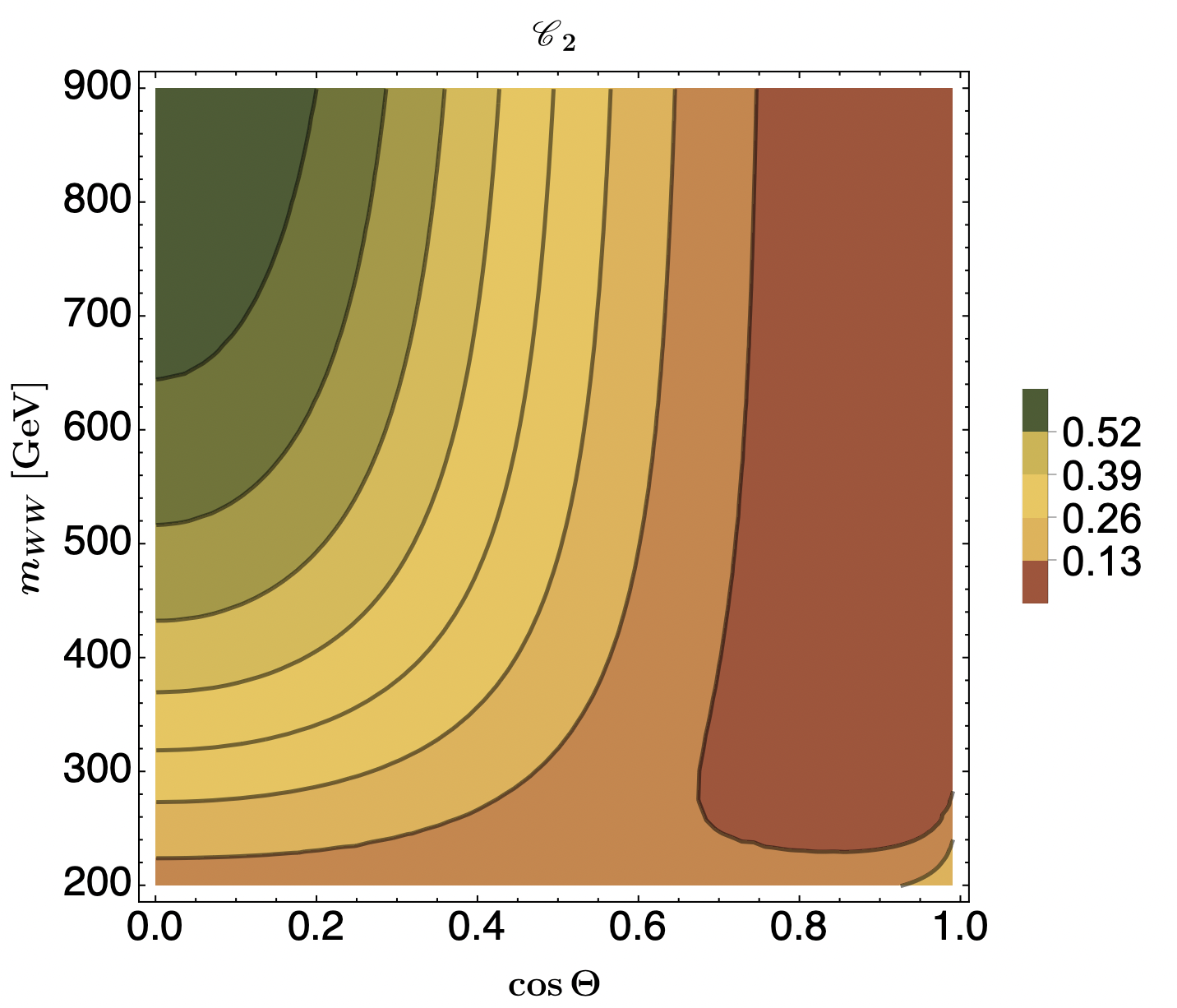}
  \includegraphics[width=3in]{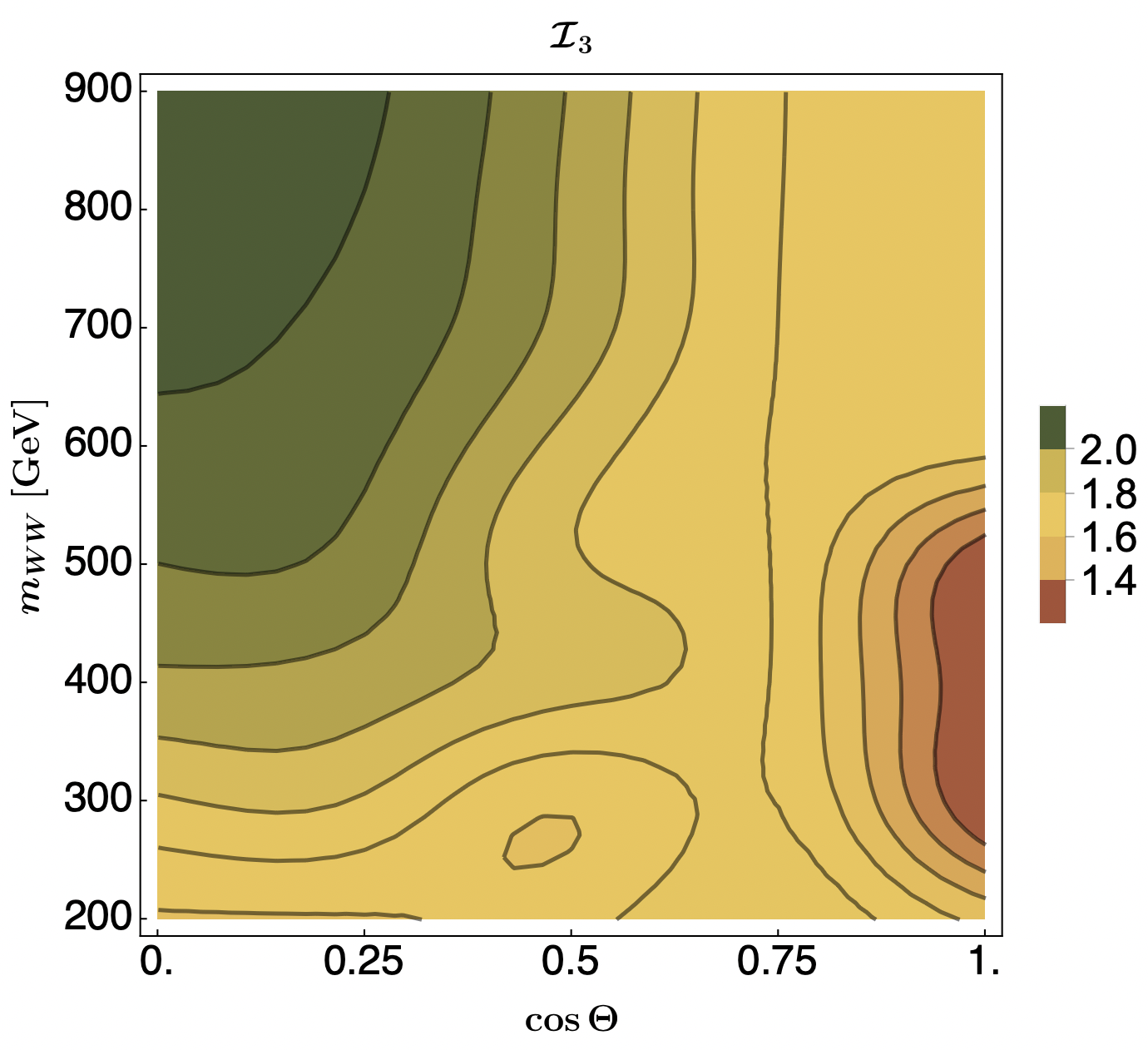}
  \caption{\small 
    The observables $\mathscr{C}_2$ (left) and ${\cal I}_3$ (right) for the process $pp \to W^+ W^-$ are shown as functions of the invariant mass and scattering angle. (Figures from \cite{Barr:2024djo} and adapted from \cite{Fabbrichesi:2023cev} under the \href{https://creativecommons.org/licenses/by/4.0/}{CC BY 4.0} license.)}
  \label{fig:WW} 
  \end{center}
  \end{figure}

 \begin{figure}[h!]
\begin{center}
\includegraphics[width=3.1in]{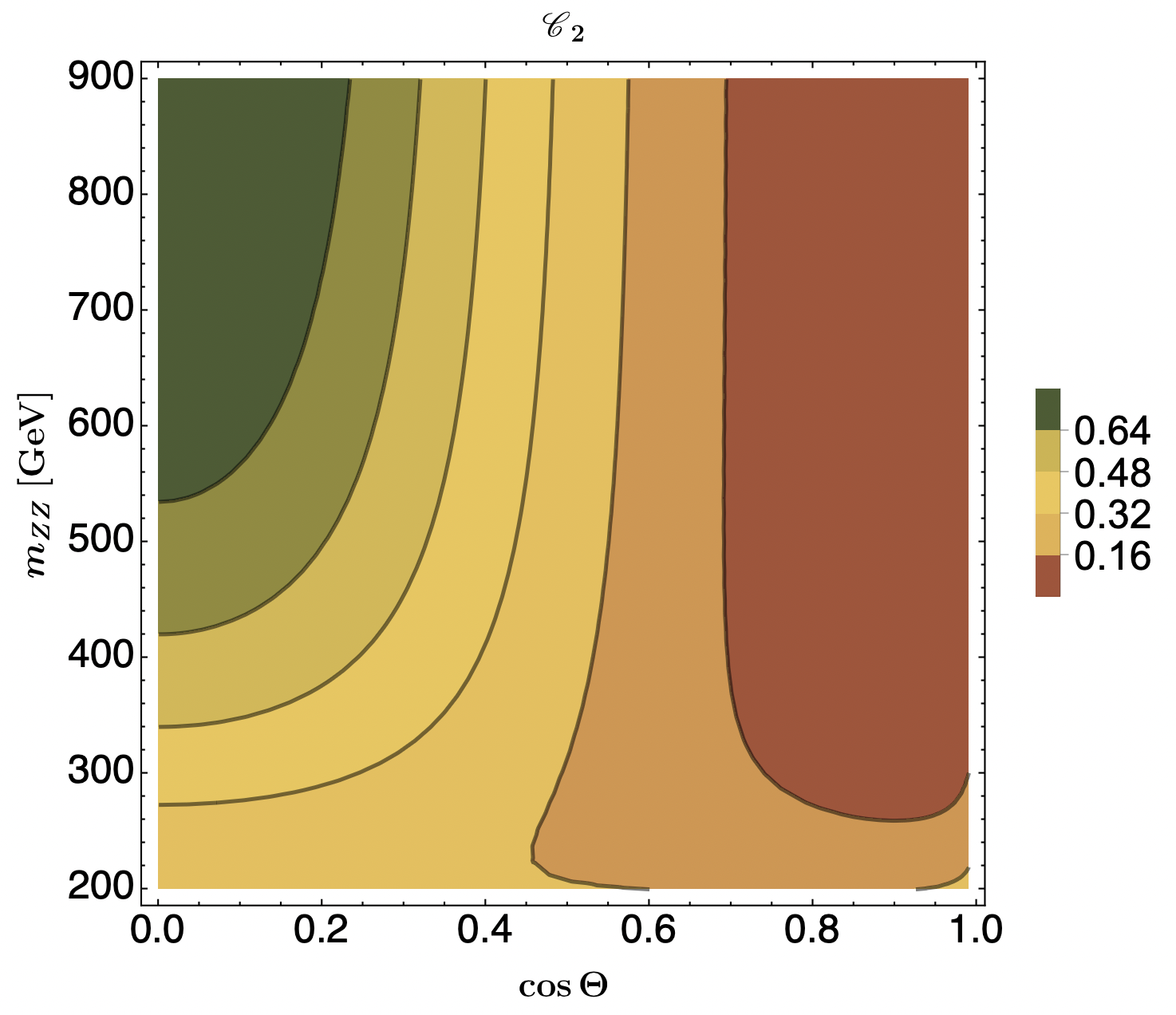}
\includegraphics[width=3in]{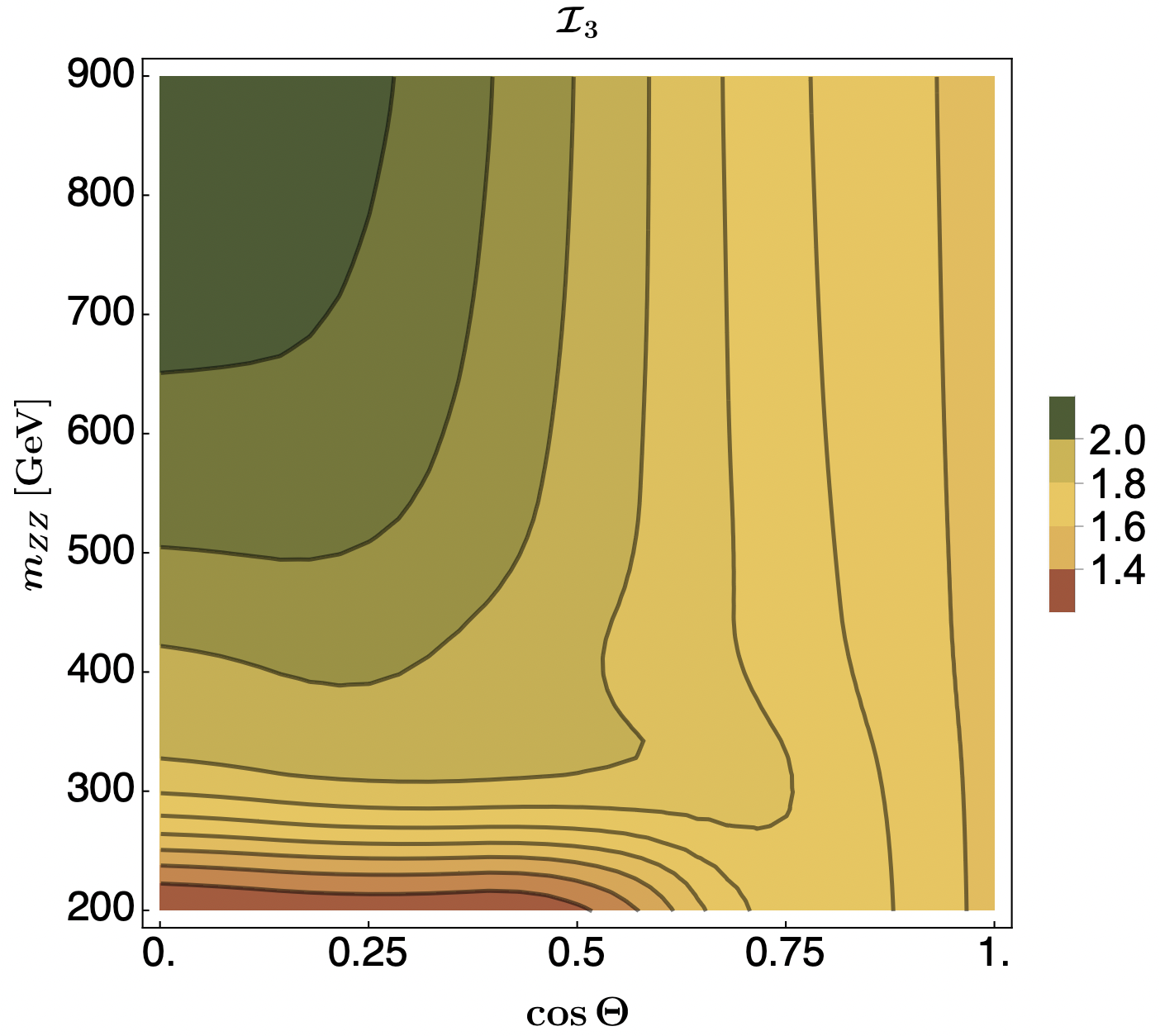}
\caption{\small
The observables $\mathscr{C}_2$ (left) and ${\cal I}_3$ (right) for the process $pp \to Z Z$ are shown as functions of the invariant mass and scattering angle.(Figures  from \cite{Barr:2024djo} and adapted from \cite{Fabbrichesi:2023cev} under the \href{https://creativecommons.org/licenses/by/4.0/}{CC BY 4.0} license.) }
\label{fig:ZZ} 
\end{center}
\end{figure}

The same study for the $W^+W^-$ and $ZZ$ production at $e^+e^-$ and muon colliders has been performed in \cite{Fabbrichesi:2023cev} via the process
 \bea
\ell^+ \ell^-\to W^+ W^-\, ,~~ZZ\, ,
 \eea
 where $\ell=e,\mu$.  In this case, due to the absence of PDF, the correlation coefficients $h_{ab}$  are given by
\bea
h_{ab} [\mVV,\Theta]&=& \frac{
\htee_{ab}[\mVV,\Theta]}{\Aee[\mVV,\Theta]}\, ,
\label{rholepton}
\eea
and analogously for the polarization coefficients $f_{a}$, $g_{a}$.
In the case of the $W^+W^-$ production the scattering angle $\Theta$ in
\eq{rholepton} is defined as the angle between the anti-lepton and $W^+$ momenta. Final lepton states were considered in the $WW$ and $ZZ$ channels, along with semi-leptonic $W$ decays.

Regarding the $W^+W^-$ production, in \cite{Fabbrichesi:2023cev} it was shown that the violation of the Bell inequalities takes place in a  range of the kinematic variables broader than in the LHC case and it is larger, while the theoretical uncertainty is negligible. On the other hand, in the case of $ZZ$ the violation of the Bell inequalities in this case takes place in a  range of the kinematic variables more or less equivalent to that at the LHC.
These analyses were based on the relevant cross sections calculated at leading order using MADGRAPH5 and with the resulting event counts scaled by the lepton identification efficiency, assumed--conservatively--to be 70\% per lepton, as in the LHC case. 

Considering the benchmark center of mass energies of a lepton (and muon) collider at $\sqrt{s}=1{\rm TeV}$ and a Future Circular Collider (FCC-ee)~\cite{Bernardi:2022hny} at $\sqrt{s}=368{\rm GeV}$,   in~\cite{Fabbrichesi:2023cev} it was shown that  for WW di-bosons, both the future muon collider~\cite{Delahaye:2019omf}, and the FCC-ee achieve a significance of 2 in rejecting the null hypothesis ${\cal I}_3\le 2$. On the other hand, for the $ZZ$ di-bosons production, the muon collider reaches the same $2\sigma$ significance, while the FCC--thanks to its larger event yield--can surpass $4\sigma$ significance~\cite{Fabbrichesi:2023cev}.

\subsubsection{Higgs decay into $WW^*$ and $ZZ^*$}
We consider here the two-qutrit system arising from weak diboson production at the LHC through resonant Higgs $(h)$ decays, produced as an $s$-channel resonance:
\be
h \to V(k_1, \lambda_1)\, V^*(k_2, \lambda_2)\, ,
\label{HVV}
\ee
where $V \in \{W, Z\}$ and $V^*$ denotes an off-shell vector boson.

The first study of this system was carried out in \cite{Barr:2021zcp}, where entanglement and the violation of Bell inequalities were investigated in the decay $h \to WW^*$. The polarization density matrix was reconstructed from the angular distributions of the charged leptons in the $W$ decays, using Monte Carlo simulations.

Subsequent analyses of the same channel appeared in \cite{Aguilar-Saavedra:2022mpg,Bi:2023uop,Fabbri:2023ncz}. The investigation was then extended to Higgs decays into two neutral gauge bosons, $ZZ$, first in \cite{Aguilar-Saavedra:2022wam,Ashby-Pickering:2022umy} and later in \cite{Bernal:2023ruk}. Comparable results for both $WW^*$ and $ZZ^*$ production were obtained in \cite{Fabbrichesi:2023cev}, where analytical expressions for the polarization density matrix of the two gauge bosons in the helicity basis were employed. 
In what follows, we summarize the main analytical results from the SM predictions as presented in \cite{Fabbrichesi:2023cev}.

For convenience, the off-shell boson can be approximated as an on-shell particle with an effective mass given by
\be
M_{V^*} = f M_V, ,
\ee
where $M_V$ is the physical mass and $f$ is a reduction factor in the range $0 < f < 1$. Following the method outlined in section \ref{sec:helicity-amplitude}, the polarization density matrix $\rho$ of the two vector bosons produced in resonant Higgs decay can be expressed in the helicity basis as~\cite{Aguilar-Saavedra:2022wam}
\be
\rho  = |\Psi \rangle \langle \Psi | \, ,
\ee
where (in the basis $|\lambda\, \lambdap\rangle = |\lambda\rangle\otimes|\lambdap\rangle$ with $\lambda,\lambdap\in\{+,0,-\}$)  
\be
|\Psi  \rangle = \frac{1}{\sqrt{2 +  \varkappa^2}} \left[  |{\small +-}\rangle -  \varkappa \,|{\small 0\, 0}\rangle + |{\small -+}\rangle  \right] \label{pure}
\ee
with
\be
 \varkappa = 1+ \frac{m_h^2 - (1+f)^2 M_V^2}{2 f M^2_V} 
\ee
and $ \varkappa=1$ corresponding to the production of two gauge bosons at rest.
In the Gell-Mann basis this is given by~\cite{Fabbrichesi:2023cev}.
\be
\rho = 2 \begin{pmatrix} 
  0 & 0 & 0 & 0 & 0 & 0 & 0 & 0 & 0  \\
  0 & 0 & 0 & 0 & 0 & 0 & 0 & 0 & 0  \\
  0 & 0 &  h_{44} & 0 &  h_{16} & 0 & h_{44} & 0 & 0  \\
  0 & 0 & 0 & 0 & 0 & 0 & 0 & 0 & 0  \\
  0 & 0 &  h_{16} & 0 & 2\, h_{33} & 0 & h_{16} & 0 & 0  \\
  0 & 0 & 0 & 0 & 0 & 0 & 0 & 0 & 0  \\
  0 & 0 &  h_{44} & 0 &  h_{16} & 0 &  h_{44}& 0 & 0  \\
  0 & 0 & 0 & 0 & 0 & 0 & 0 & 0 & 0  \\
  0 & 0 & 0 & 0 & 0 & 0 & 0 & 0 & 0  \\
\end{pmatrix} \, ,
\label{rhoH}
\ee
with the condition $\Tr[\rho_H]=1$ following from the relation $4(h_{33}+h_{44})=1$. 
Under these assumptions, only two independent coefficients remain.
Analytical expressions for the $h_{ab}$ in \eq{rhoH} can be found in~\cite{Fabbrichesi:2023cev}.
Same spin density matrix as in \eq{rhoH} has been computed in terms of tensor components $T^L_M$  in \cite{Aguilar-Saavedra:2022wam}, whose correlation coefficients are related to those in  \eq{rhoH} by 
\be
\frac{1}{6} C_{2,2,2,-2}= h_{44}  \quad \text{and}  \quad \frac{1}{6} C_{2,1,2,-1} =h_{16}  \, . \label{Ctensors}
\ee
where the coefficients $C_{2,2,2,-2},~C_{2,1,2,-1}$ are defined in \eq{eq:rho-tensor2}.

The density matrix in \eq{rhoH} represents a pure state, as can be verified from the condition $\rho^2=\rho$.  It should be emphasized, however, that this purity holds only for a fixed value of the invariant mass of the off-shell gauge boson. Then, under the assumption that the diboson system is described by a pure state, then one can measure its entanglement through the  entropy of entanglement defined in \eq{entropy}.

Figures~\ref{fig:HWW} and \ref{fig:HZZ} show the entanglement entropy ${\cal E}$ and the maximized Bell operator value ${\cal I}_3$ for $WW^*$ and $ZZ^*$ decays~\cite{Fabbrichesi:2023cev} respectively. The maximization of ${\cal I}_3$, dependent only on $M_V^{*}$, is performed point by point via the unitary rotation of Eq.\eq{uni_rot}. The unitary matrices yielding the maximum in the last bins ($M_{W^*}=40$ GeV, $M_{Z^*}=32$ GeV) are given in\cite{Fabbrichesi:2023cev}.

\begin{figure}[h!]
\begin{center}
\includegraphics[width=3in]{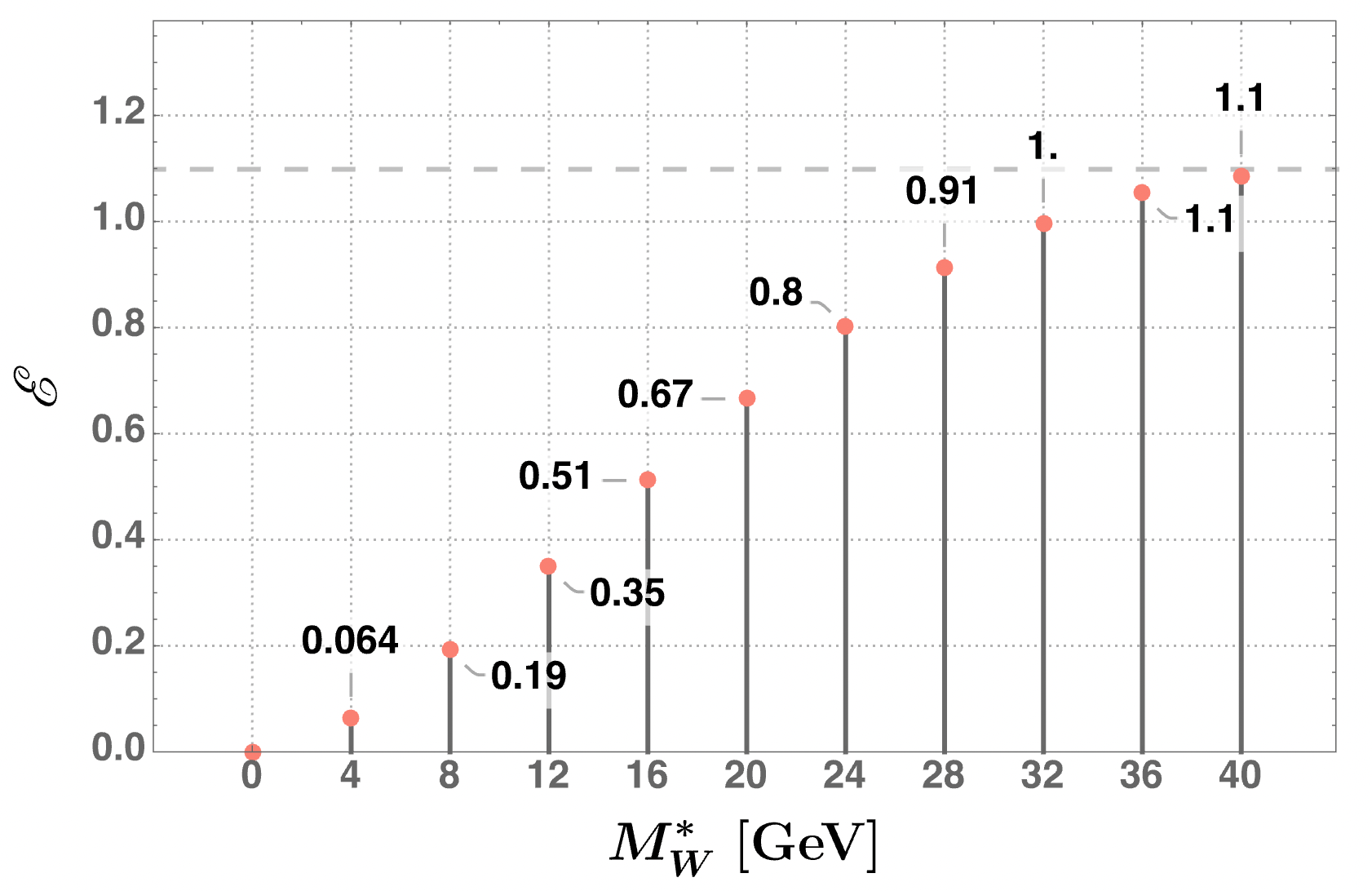}
\includegraphics[width=3in]{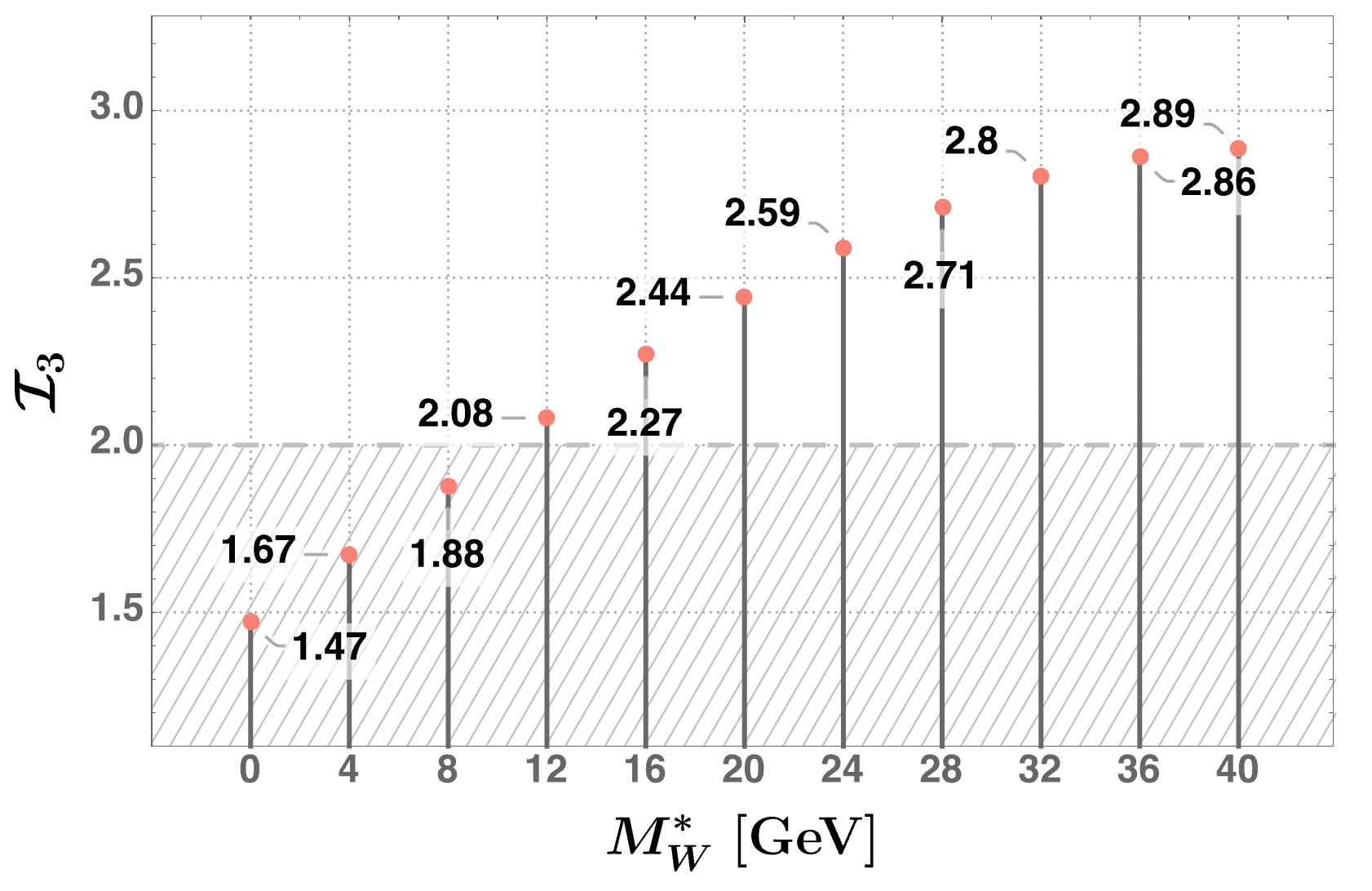}
\caption{\small  Predictions of the entanglement entropy ${\cal E}$ (left) and Bell operator value ${\cal I}3$ (right) in Higgs decays to $WW^*$, as functions of the virtual mass $0<M_{W^*}<40$GeV\cite{Fabbrichesi:2023cev}. The dashed lines indicate $\ln 3$ (left) and the Bell-violation threshold ${\cal I}_3>2$ (right). (Figures from \cite{Barr:2024djo} and adapted from \cite{Fabbrichesi:2023cev} under the \href{https://creativecommons.org/licenses/by/4.0/}{CC BY 4.0} licence).
\label{fig:HWW}
}
\end{center}
\end{figure}

\begin{figure}[h!]
\begin{center}
\includegraphics[width=3in]{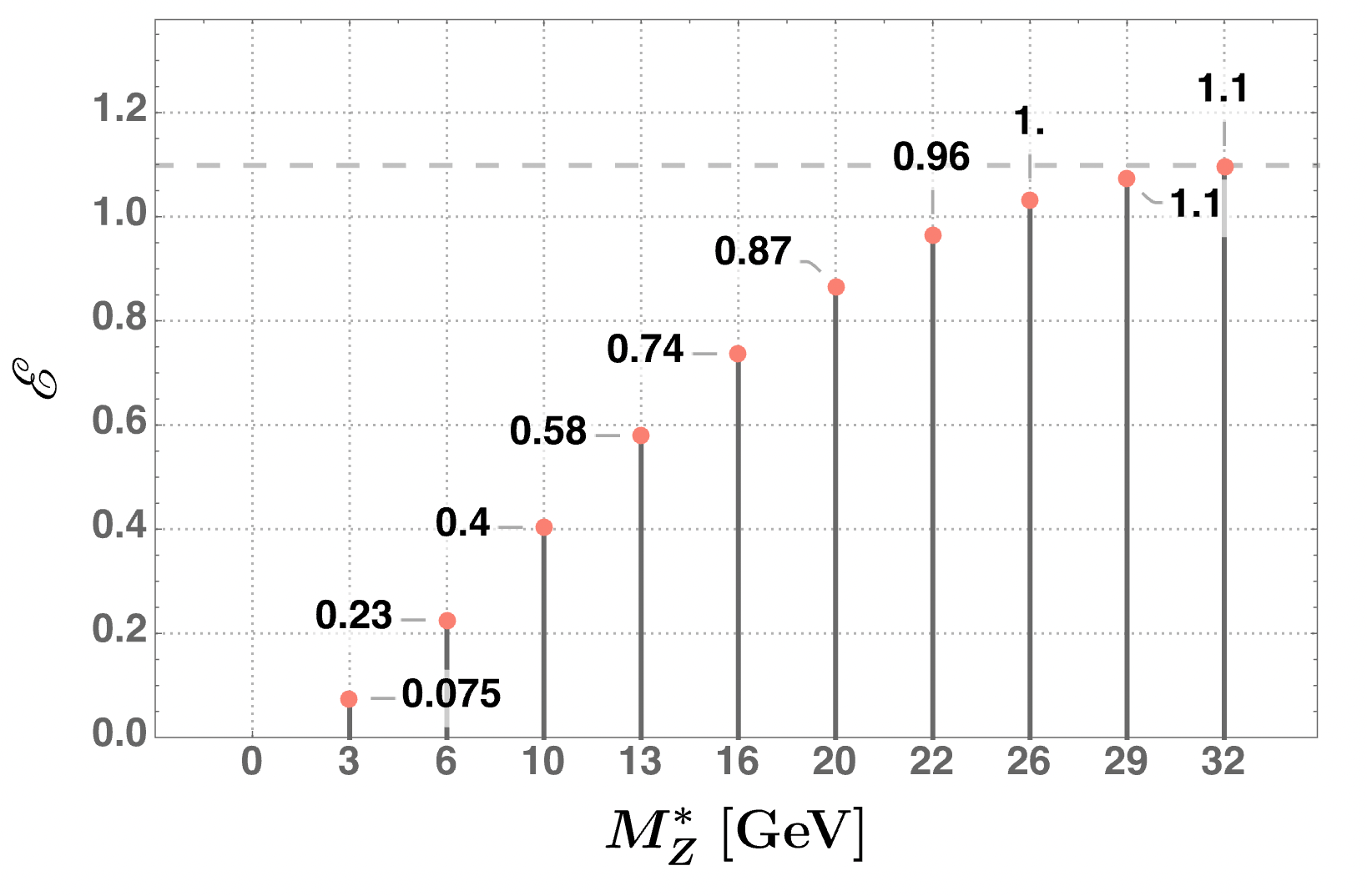}
\includegraphics[width=3in]{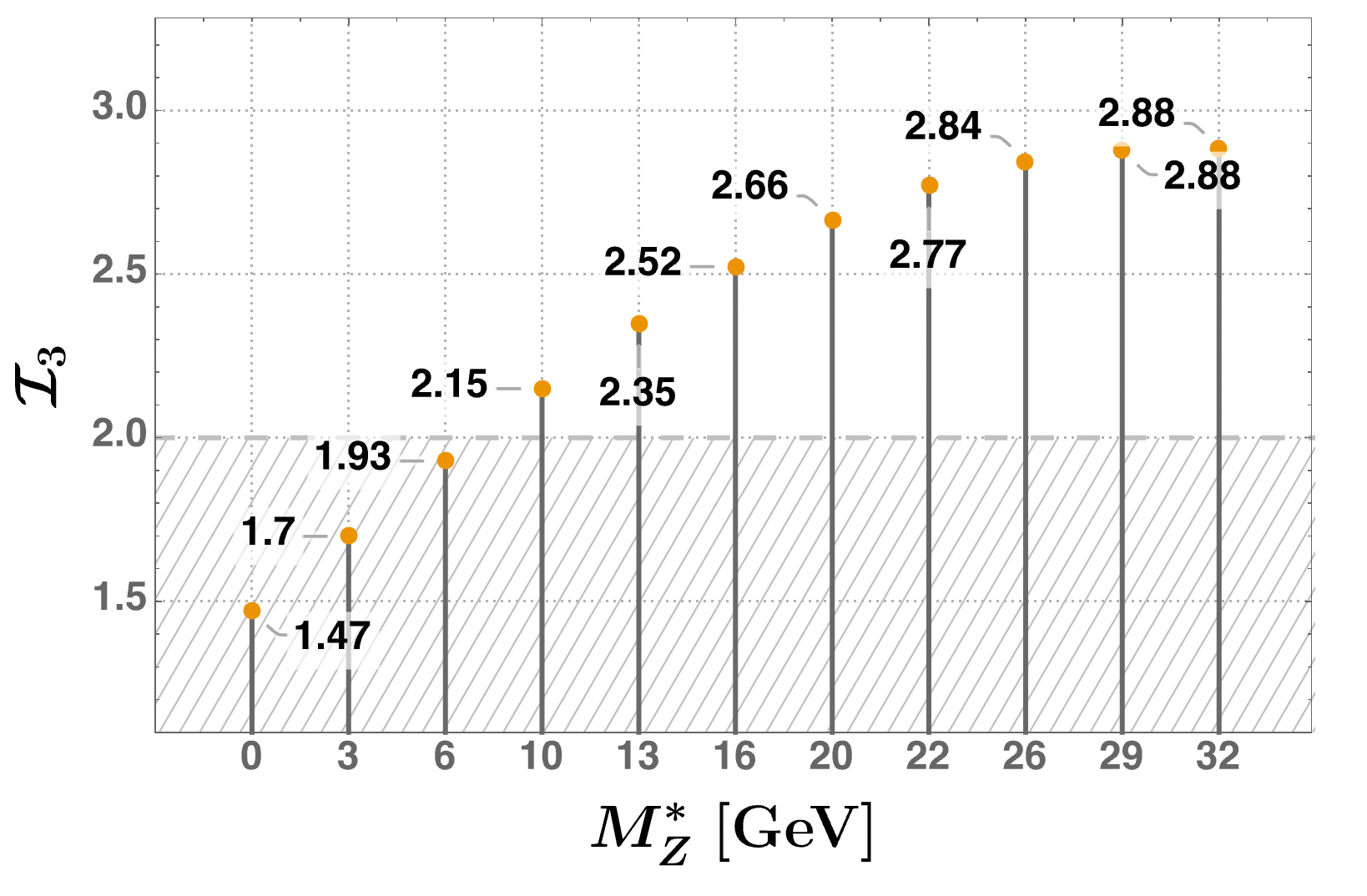}
\caption{\small Same as in Fig.\ref{fig:HWW} but 
  for the pair production of $Z$ bosons in Higgs boson decays as functions of the virtual $Z^{*}$  mass in the range $0<M_{Z^*}<32$~GeV~\cite{Fabbrichesi:2023cev}. (Figures from \cite{Barr:2024djo} and adapted from \cite{Fabbrichesi:2023cev} under the \href{https://creativecommons.org/licenses/by/4.0/}{CC BY 4.0} licence). \label{fig:HZZ} 
}
\end{center}
\end{figure}
The process $h\to WW^*$ was first simulated in~\cite{Barr:2021zcp} using
\textsc{MadGraph5\_aMC@NLO} with full spin correlations and relativistic effects. Only leptonic decays were considered, with the Bell operator ${\cal I}_3$ optimized in Cartesian planes. Due to two neutrinos in the final state, rest-frame reconstruction is uncertain: depending on momentum smearing, the significance of Bell violation ranges from about $5\sigma$ to $1\sigma$ at 140 fb$^{-1}$. Semi-leptonic decays $h \to jj \ell \nu\ell$ were later studied in\cite{Fabbri:2023ncz}, where $c$-tagging reduces uncertainties, though the large $WW^*$ background remains a limitation.

For $h\to ZZ^*$, simulations with tensor~\cite{Aguilar-Saavedra:2022wam} and Gell-Mann~\cite{Ashby-Pickering:2022umy} bases show a much cleaner situation: no neutrinos, precise rest-frame reconstruction, and low background. Using
\textsc{MadGraph5\_aMC@NLO}, the Bell operator basis is explicitly determined, and a significance up to $4.5\sigma$ is expected at 3 ab$^{-1}$, making this channel the most promising test of Bell inequality in weak boson decays. All the $WW^*$ and $ZZ^*$ results based on Monte Carlo simulation are consistent with the significance estimated in ~\cite{Fabbrichesi:2023cev}.

\subsubsection{Vector-boson fusion}
We conclude this section by summarizing the study of di-boson production via vector-boson fusion presented in~\cite{Morales:2023gow},
\be
W^+W^- \to W^+W^-,, \quad Z\gamma \to W^+W^- ,, \quad \gamma\gamma \to W^+W^- \, ,
\ee
where the corresponding tree-level SM amplitudes were computed. This mechanism gives rise to different final-state configurations: two qubits (photons), one qubit and one qutrit (a photon with a massive gauge boson), or two qutrits (massive gauge bosons). The analysis focused on both entanglement and Bell inequality violation. While the degree of entanglement depends on the phase-space region, most channels display a similar level of entanglement, with the exception of $ZZ\to ZZ$, where it is strongly suppressed. Bell inequality tests can be carried out in this framework by evaluating the expectation value of the appropriate Bell operator in selected phase-space regions, as detailed in~\cite{Morales:2023gow}.

\subsubsection{Probing new physics with entanglement}
The potential of entanglement and quantum observables to probe non-standard Higgs couplings to massive gauge bosons has been studied in~\cite{Fabbrichesi:2023jep,Bernal:2023ruk}, where the most general Lorentz-invariant interaction Lagrangian between the Higgs boson $h$ and the gauge bosons $W^\pm$ and $Z$ is considered (see equation below)
\begin{align}
{\cal L}_{hVV}  = & g\, M_{W} W^{+}_{\mu}W^{-\mu} h + \frac{g}{2 \cos\theta_{W}} M_{Z} Z_{\mu}Z^{\mu} h\nn \\
& -\frac{g}{M_{W} }\Bigg[ \frac{a_{W}}{2} W^{+}_{\mu\nu} W^{-\mu\nu} +
\frac{\widetilde a_{W}}{2} W^{+}_{\mu\nu}\widetilde W^{-\mu\nu}  +\,\frac{a_{Z}}{4} Z_{\mu\nu} Z^{\mu\nu}  + 
\frac{\widetilde a_{Z}}{4} Z_{\mu\nu}\widetilde Z^{\mu\nu}\Bigg]  h\, ,\label{eq:Lhvv}\, ,
\end{align}
where $V^{\mu\nu}$ is the field strength tensor of the gauge boson $V=W$ or $Z$ and $\widetilde V^{\mu\nu}=\frac12 \epsilon^{\mu\nu\rho\sigma}V_{\rho\sigma}$ is the corresponding dual tensor. The anomalous couplings $\widetilde a_{V}$ signals  the violation of the CP symmetry through the interference with the SM contribution.

Constraints on these couplings were studied in~\cite{Fabbrichesi:2023jep} using an observable sensitive to the anti-symmetric part of the density matrix
 $\mathscr{C}_{odd}=\frac{1}{2}\, \sum_{\substack{a,b\\ a< b}} \Big| h_{ab} -h_{ba} \Big|$ and entropy of entanglement defined in Eq.~\eqref{entropy}. A $\chi^2$ test at 95\% CL was performed, with uncertainties estimated from the Higgs mass measurements in $pp \to h \to W\ell^-\bar{\nu}$\cite{ATLAS:2022ooq} and $pp \to h \to Z\ell^+\ell^-$\cite{ATLAS:2020rej}. These errors, taken as proxies for the uncertainty in reconstructing the Higgs rest frame and hence gauge boson polarizations, were propagated to the observables via a Monte Carlo varying $m_h$ within experimental limits. The main results of this analysis are summarized in Table~\ref{tab:anomalous-Higgss} for the LHC-run2 and Hi-Lumi benchmarks luminosities of ${\cal L}=140~{\rm fb}^{-1}$ and ${\cal L} =3~{\rm ab}^{-1}$ respectively.

\begin{table}[t]
\centering
\begin{tabular}{*{2}{p{0.15\textwidth}}}
  \toprule
       LHC run2&
       LHC Hi-Lumi  \\
  \midrule
   $ |a_{W}| \leq 0.033$    & $|a_{W}|\leq 0.0070 $ \\
     $|\widetilde a_{W}|\leq 0.031$   & $|\widetilde a_{W}|\leq 0.0068$  \\
   $ |a_{Z}| \leq 0.0019 $   & $|a_{Z}|\leq 0.00040$ \\
     $|\widetilde a_{Z}|\leq 0.0039$   & $|\widetilde a_{Z}|\leq 0.00086$  \\
  \bottomrule%
\end{tabular}
\caption{95\% joint confidence intervals for the anomalous couplings from LHC data (backgrounds neglected), using the entropy of entanglement and $\mathscr{C}_{odd}$ in the $\chi^2$ test~\cite{Fabbrichesi:2023jep}, for the LHC run2 (${\cal L}=140~{\rm fb}^{-1}$) and Hi-Lumi (${\cal L} =3~{\rm ab}^{-1}$) scenarios.}
\label{tab:anomalous-Higgss}
\end{table}


\subsection{Tau-lepton pair production}
The study of entanglement in $\tau$-lepton pairs was first proposed at LEP~\cite{Privitera:1991nz}, and later extended to the LHC~\cite{Fabbrichesi:2022ovb} and SuperKEKB~\cite{Ehataht:2023zzt}. At the LHC, the polarization density matrix is computed analogously to the top-quark case of Section~\ref{sec:top}, with the main production mechanism being Drell–Yan: quark annihilation into a photon or $Z$ boson in the $s$-channel, followed by decay into a $\tau$-pair. The corresponding tree-level diagrams are obtained from the third diagram in Fig.~\ref{fig:ttbar}  by replacing the propagator in the $s$-channel gluon with a photon or $Z$ boson.

For the $\tau$ pair production process at the LHC
\be
p+p \to \tau^- + \tau^+  \, ,
\ee
the corresponding unpolarized cross section  is given by \cite{Fabbrichesi:2022ovb}
\be
\frac{\di \sigma}{\di \Omega \, \di m_{\tau\bar\tau}} = \frac{\alpha^2 \beta_\tau}{64 \pi^2 m_{\tau \bar \tau}^2} \Big\{ L^{uu} (\tau) \, \tilde A^{uu}[m_{\tau\bar\tau},\, \Theta]+\big[L^{dd} (\tau)+L^{ss}(\tau) \big]\, \tilde A^{dd}[m_{\tau\bar\tau},\, \Theta]  \Big\}\, ,
\ee
where $\tau = m_{\tau^- \tau^+}/\sqrt{s}$,
$L^{uu,dd,ss}(\tau)$ the luminosity functions as defined in \eq{eq:lumPDF}, and the electromagnetic fine-structure constant $\alpha=e^2/4 \pi$. The correlation spin coefficients $C_{ij}$--expressed in the basis of \eq{basis-rnk}--can be obtained by following the method explained in the $t\bar{t}$ production in section \ref{sec:top}, and are given by \cite{Fabbrichesi:2022ovb}
\be
C_{ij} [m_{t\bar t},\, \Theta]= \frac{L^{uu} (\tau) \, \tilde C_{ij}^{uu}[m_{\tau\bar\tau},\, \Theta]+\big[L^{dd} (\tau)+L^{ss} (\tau) \big]\, \tilde C_{ij}^{dd}[m_{\tau\bar\tau},\, \Theta]} {L^{uu}(\tau) \, \tilde A^{uu}[m_{\tau\bar\tau},\, \Theta]+\big[L^{dd} (\tau)+L^{ss} (\tau)\big] \, \tilde A^{dd}[m_{\tau\bar\tau},\, \Theta]} \label{eq:Cij-tau}\, ,
\ee
where the down-quark luminosity functions can be grouped together because they multiply the same correlation functions. The analytical expressions for the 
$C_{ij}^{uu,dd,ss}[m_{\tau\bar\tau},\, \Theta]$ functions can be found in \cite{Fabbrichesi:2022ovb}, where the $\Theta$ scattering angle is defined as the angle between the initial quark and $\tau^-$ momenta in the $\tau^+\tau^-$ center of mass frame.

A much simpler formula holds at lepton colliders for the process
\be
e^+ + e^- \to \tau^- + \tau^+  \, .
\label{eq:ee}
\ee
because there are no PDF luminosity functions and  the CM energy is fixed.
Results for the concurrence and Bell inequality violations are shown in Fig.\ref{fig:m1m2tau}.

 \begin{figure}[h!]
\begin{center}
\includegraphics[width=3in]{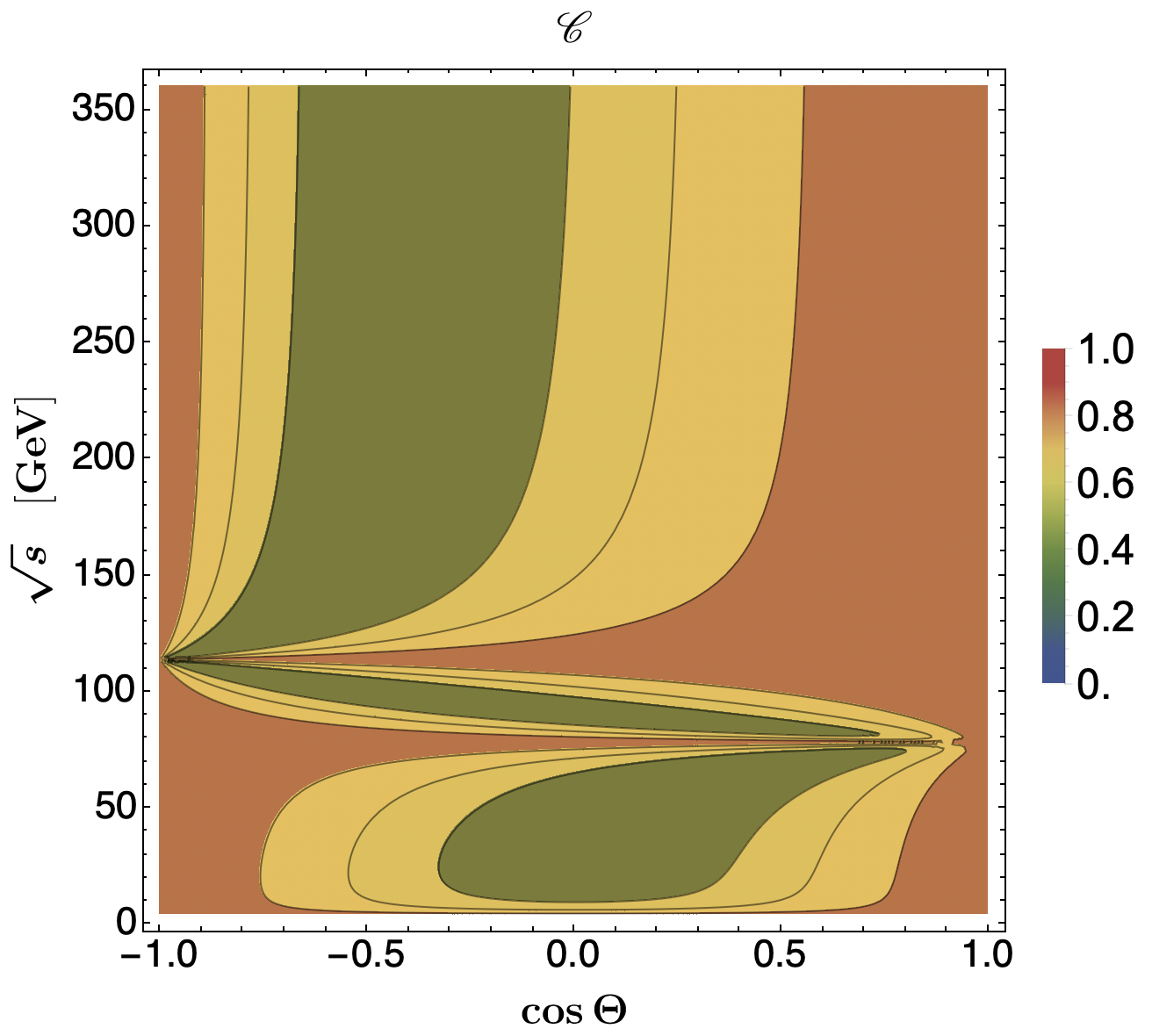}
\includegraphics[width=3in]{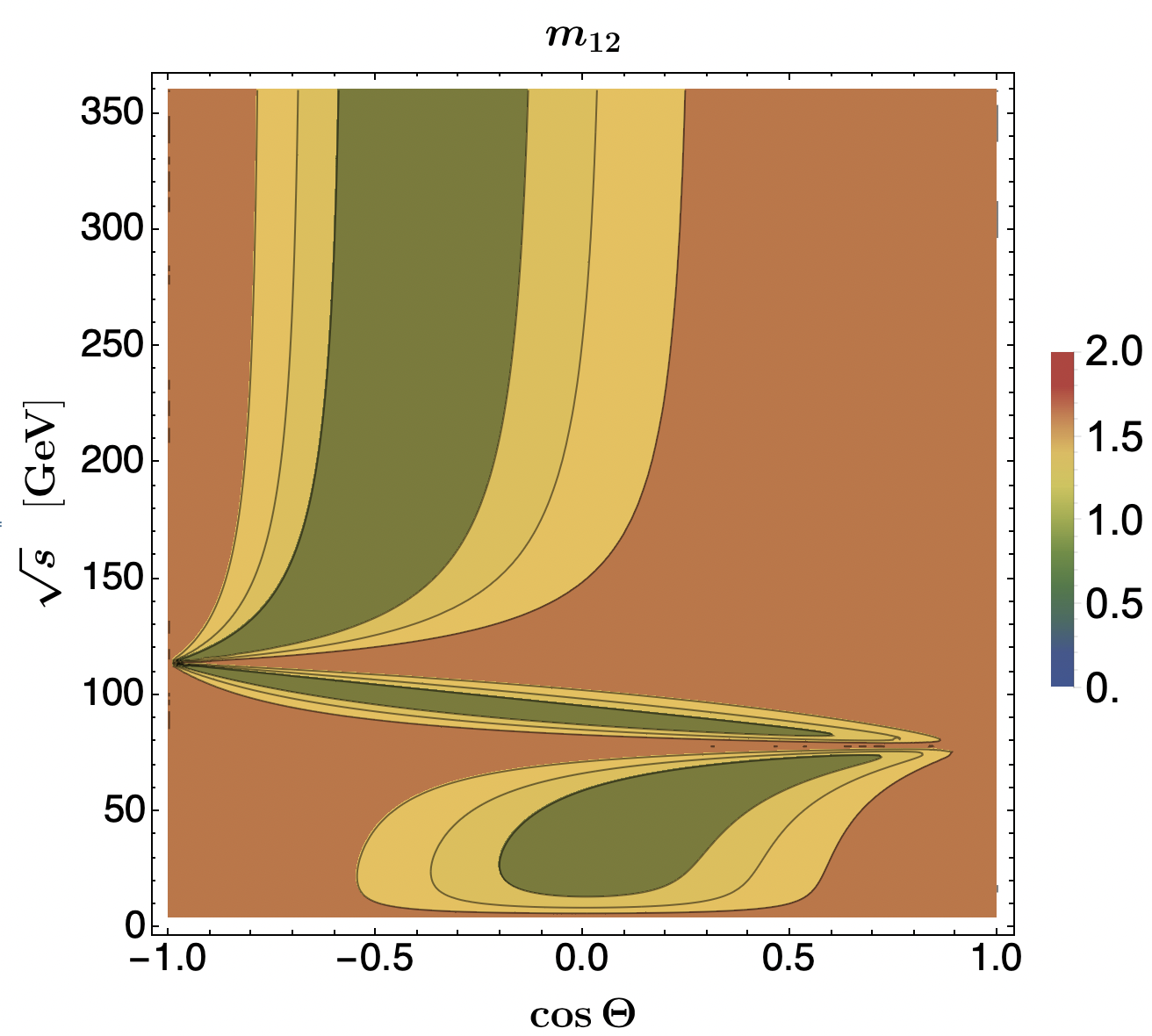}
\caption{\small Concurrence ($\mathscr{C}$) and  $\mathfrak{m}_{12}$ for the $e^+ e^- \to \tau \bar{\tau}$ pair production, as a function of the kinematic variables $\Theta$ and $m_{\tau \bar \tau}$ across the entire available space.
(Figures from \cite{Barr:2024djo} and adapted from \cite{Fabbrichesi:2022ovb}
under the \href{https://creativecommons.org/licenses/by/4.0/}{CC BY 4.0} licence).}
\label{fig:m1m2tau} 

\end{center}
\end{figure}
 
At SuperKEKB, operating at a center-of-mass energy of $\sqrt{s} = 10 ,\text{GeV}$, $\tau$-pair production is dominated by the single-photon exchange diagram. The quantum entanglement properties and Bell inequality violations in $\tau$-pair production at SuperKEKB have been investigated in~\cite{Ehataht:2023zzt}. It was shown that $\tau$-pair production at SuperKEKB provides an excellent environment for studying entanglement and Bell inequality violation, owing to the large event sample and the exceptionally clean experimental conditions. In this case, both the expected concurrence $\mathscr{C}$ and the Bell-inequality correlator $\mathfrak{m}_{12}$ can be expressed in compact analytic form~\cite{Ehataht:2023zzt}:
\bea
    \mathscr{C}=\frac{\left(s - 4 \,m_{\tau}^2\right) \sin^2\Theta}{4\, m_{\tau}^{2} \sin^2\Theta + s \left(\cos^2\Theta + 1\right)}\, , ~~
   \mathfrak{m}_{12}= 1 + \left(\frac{\left(s - 4\, m_{\tau}^2\right) \sin^2\Theta}{4 \, m_{\tau}^2 \sin^2\Theta+ s \left(\cos^2\Theta + 1\right)}\right)^{2} \,.
\eea
The maximum value for both $\mathscr{C}[\rho]$ and $\mathfrak{m}_{12}$ at SuperKEK are reached for  scattering angles close to $\pi/2$, as shown in Fig.~\ref{fig:m1m2tau}.

In \cite{Ehataht:2023zzt}, $200$ million $e^+e^- \to \tau^+\tau^-$ events were simulated with \textsc{MadGraph5\_aMC@NLO}~\cite{Alwall:2014hca} at leading order and processed with \textsc{PYTHIA}~\cite{Sjostrand:2014zea} for parton showers, hadronization, and $\tau$ decays. 
All relevant $\tau$ channels were included, and results were studied both at truth level and with detector effects. 
The analysis predicts entanglement and Bell inequality violation with $>5\sigma$ significance, for datasets comparable to Belle~II.

Quantum entanglement and Bell inequality violations in the tau pair has also been investigated in the resonant Higgs decay process~\cite{Fabbrichesi:2022ovb,Altakach:2022ywa,Ma:2023yvd}
\bea
H\to \tau^+~\tau^-\, .
\eea
Because the final states originate from the decay a scalar particle, a pure state should be created for the spins.
From the Yukawa coupling interaction of the Higgs boson with fermions, the spin correlation matrix of the final tau-pair is \cite{Fabbrichesi:2022ovb} 
\be
C = \begin{pmatrix} 1&0&0\\ 0&1&0 \\ 0&0&-1 \end{pmatrix} ,
\ee
defined in the $\{\hat{n},\hat{r},\hat{k}\}$ basis of \eq{basis-rnk}.
This leads to maximal entanglement, ${\cal C}[\rho]=1$, and $\mathfrak{m}_{12}=2$, implying maximal violation of Bell inequality.

The decay of the Higgs boson into $\tau$-lepton pairs has been analyzed by Monte Carlo simulations in \cite{Altakach:2022ywa} and \cite{Ma:2023yvd} by means of the $HZ$ associated production at future $e^+e^-$ colliders--where it would be easier to reconstruct the final state with respect to the $s$-channel production in hadron colliders.
In \cite{Altakach:2022ywa}, entanglement is predicted to be observable above $5\sigma$ at both the International Linear Collider (ILC) \cite{Behnke:2013xla} and FCC-ee. Bell inequality violation is not expected at the ILC, but should reach about $3\sigma$ at the FCC-ee, while on the other hand for the Circular Electron Positron Collider (CEPC)~\cite{CEPCStudyGroup:2018rmc,CEPCStudyGroup:2018ghi}, the expected significance is only $\sim 1\sigma$~\cite{Ma:2023yvd}.

\subsubsection{Probing new physics with entanglement}
The sensitivity of $\tau$-pair entanglement to new physics has been explored in the framework of contact interactions between quarks and $\tau$-leptons \cite{Fabbrichesi:2022ovb} and in
CP properties of the coupling to the Higgs boson in \cite{Fabbrichesi:2022ovb,Altakach:2022ywa}.

  Regarding the contact interactions, these can be probed at the LHC.
  In \cite{Fabbrichesi:2022ovb} it is shown that entanglement is most sensitive to new-physics effects at high energies, with the optimal region identified as
  $m_{\tau\bar{\tau}}> 800~{\rm GeV} $ and scattering angle close to $\pi/2$,
where one can have enough events and new-physics effects sizable.
In this regime, the relative difference between the SM and new physics with $\Lambda=25~{\rm TeV}$ reaches $\sim 70\%$. At the Hi-Lumi LHC, the SM can be rejected with a significance of $2.7\sigma$ for such a contact interaction, in line with current bounds on four-fermion operators~\cite{ParticleDataGroup:2022pth}.

Concerning the CP properties of the Higgs boson, the authors in \cite{Altakach:2022ywa} considered  the associated production $Zh$ at $e^+e^-$ colliders and look in the subsequent decay $h\to \tau^+\tau^-$  at the generic interaction Lagragian
\be
{\cal L}_h = - \frac{m_\tau}{v} \kappa \, h \, \bar \tau \Big( \cos \delta + i \gamma_5 \sin \delta \Big) \tau \, ,
\ee
where $\delta$ is the CP violating phase.
Quantum state tomography of the decay is proposed via the computation of the correlation matrix, which-- in the basis of \eq{basis-rnk}-- is  given by
\be
C_{ij} =  \begin{pmatrix} \cos 2 \delta & \sin 2 \delta& 0 \\
-\sin 2 \delta&  \cos 2 \delta &0\\ 0 & 0 &-1\end{pmatrix}\,. 
\ee
A Monte Carlo analysis with \textsc{MadGraph5\_aMC@NLO} has been performed
in~\cite{Altakach:2022ywa} for two benchmark colliders: the ILC and FCC-ee, by using the  one-prong decays $\tau^+ \to \pi^+ \nu_\tau$ and  $\tau^- \to \pi^- \bar \nu_\tau$ to reconstruct the $\tau$-pair polarization density matrix $C_{ij}$ matrix. Constraints at 90\% C.L. were obtained for  $\delta$ to the intervals: 
$-10.89^\circ\le \delta \le 9.21^\circ~\text{(ILC)}$ and $7.36^\circ\le \delta \le 7.31^\circ~\text{(FCC-ee)}$, which are comparable to the constraints obtained by means of traditional analysis ~\cite{Berge:2013jra}.

\section{Outlook}
\label{sec:conclusions}
The study of quantum entanglement and Bell inequality violations at colliders has advanced from a theoretical investigation to an experimentally accessible domain. Recent measurements at the LHC and BESIII have provided direct evidence of entanglement in top-quark pairs, B-meson decays, and charmonium states, thereby extending quantum non-locality tests to energy scales many orders of magnitude higher than traditional optical experiments. These achievements confirm that quantum information concepts can be meaningfully embedded within the framework of high-energy physics.

Looking forward, several promising directions emerge. First, the increasing precision of collider experiments will allow systematic investigations of entanglement across a wider class of processes, including diboson production and Higgs decays. Second, the development of quantum tomography techniques tailored for collider observables will be crucial for disentangling genuine quantum correlations from background effects. Third, entanglement-sensitive observables may offer novel probes of physics beyond the SM, particularly in scenarios involving CP violation, contact interactions, or non-standard Higgs couplings.

In conclusion, the intersection of quantum information and high-energy physics is opening a new frontier: not only do collider experiments provide a unique platform for testing fundamental aspects of quantum mechanics, but they also enrich the phenomenological toolbox for exploring new physics. Continued synergy between theory, experiment, and quantum information science is expected to transform this emerging field into a central component of future collider programs.

\bibliographystyle{JHEP}
{\small
\bibliography{chapter-QE-at-colliders.bib}
}
\end{document}